\documentclass[twocolumn,nofootinbib,superscriptaddress]{revtex4-1}

\pdfoutput=1

\usepackage{slashed}

\usepackage[normalem]{ulem}

\usepackage{color}

\usepackage{epsfig}
\usepackage{epstopdf}
\usepackage{epsf}
\input{epsf.sty}
\usepackage{graphicx,amsmath,amssymb}
\usepackage{epsfig}
\allowdisplaybreaks

\def\ei{\end{itemize}}
\def\be{\begin{equation}}
\def\ee{\end{equation}}
\newcommand{\bea}{\begin{eqnarray}}
\newcommand{\eea}{\end{eqnarray}}

\usepackage{bbm,bm,amsmath,amssymb}

\usepackage{hyperref}

\def\K{{K\"{a}hler}}

\newcommand{\rf}[1]{(\ref{#1})}
\newcommand{\cN}{\mathcal{N}}
\def\E{{$E_{7(7)}$}}
\def\K{K{\"a}hler}

\newcommand{\vp}{\varphi}

\usepackage{float}

\begin{document}

 \title{\Large\rm {\bf   \boldmath  CMB targets after the latest Planck data release}}

\author{Renata Kallosh}
\email{kallosh@stanford.edu}
\author{Andrei Linde}
 \email{alinde@stanford.edu}
\affiliation{Stanford Institute for Theoretical Physics and Department of Physics, Stanford University, Stanford,
CA 94305, USA}

 \begin{abstract}
We show that a combination of the simplest $\alpha$-attractors and KKLTI models related to  Dp-brane inflation covers most of the area in the ($n_{s}$, $r$) space favored by Planck 2018. For $\alpha$-attractor models, there are discrete targets $3\alpha=1,2,...,7$, predicting 7 different values of $r = 12\alpha/N^{2}$ in the range $10^{-2} \gtrsim r \gtrsim 10^{-3}$. In the small $r$ limit, $\alpha$-attractors  and Dp-brane inflation models describe  vertical $\beta$-stripes in the ($n_{s}$, $r$) space, with $n_{s}=1-\beta/N$,  $\beta=2, {5\over 3},{8\over 5}, {3\over 2},{4\over 3}$. A phenomenological description of these models and their generalizations can be achieved in the context of pole inflation.  Most of the $1\sigma$ area in the ($n_{s}$, $r$) space favored by Planck 2018 can be covered models with $\beta = 2$ and $\beta = 5/3$. 
Future precision data on $n_s$ may help to discriminate between these models even if the precision of the measurement of $r$ is insufficient for the discovery of gravitational waves produced during inflation. 
  
 \end{abstract}

\maketitle



\parskip 7pt 
 
\section{Introduction} 
 
 ~~~~~Current and future CMB missions, such as BICEP2/Keck \cite{Hui:2018cvg,Ade:2018gkx}, CMB-S4 \cite{Abazajian:2016yjj,Shandera:2019ufi,Abazajian:2019eic}, SO~\cite{Ade:2018sbj}, 
  LiteBIRD  \cite{Hazumi:2019lys} and PICO \cite{Hanany:2019lle},  may potentially  detect the tensor to scalar ratio at a level $r= 5\cdot 10^{-4} (5\sigma) $ and improve constraints on $n_s$ by a factor of three relative to Planck, to achieve $\sigma (n_s)= 0.0015$ \cite{Hanany:2019lle}.  A thorough investigation of   all phenomenologically viable inflationary models that  can explain the future CMB data  is necessary for a correct interpretation of the meaning of a detection/non-detection of the  primordial gravitational waves. It is therefore  important to perform a careful investigation of the motivation, phenomenological consistency, and  
predictions of such models. 
\begin{figure}[!h]
\hspace{-10mm} 
\includegraphics[width=9.55cm]{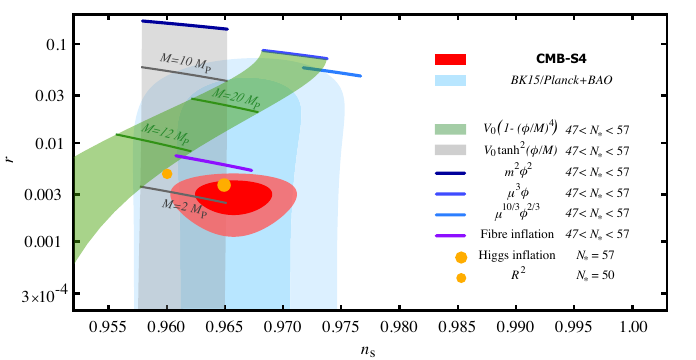}
\caption{\footnotesize The  figure from  the latest CMB-S4 Science Case paper \cite{Abazajian:2019eic}. The gray area shows predictions of the simplest $\alpha$-attractor model $V\sim \tanh^{2}  {\varphi\over M}$ for $47< N < 57$. The green area is for the hilltop  model with $V\sim 1-({\varphi/M})^4$. This  model is theoretically inconsistent for $M\gg1$, which is the only range of $M$ where it could match observational data   \cite{Kallosh:2019jnl}. }
\label{CMBS42019}
\vspace{-0.2cm}
\end{figure}

Of course, one may argue that it is premature to plan for the long journey when the goal is nearby, and the B-mode detection at $r\gtrsim 10^{-2}$ is possible. For example, power-law axion monodromy potentials  during inflation have potentials proportional to $\varphi^p$ with $p<2$  \cite{Silverstein:2008sg,McAllister:2008hb,McAllister:2014mpa}. These potentials were derived in string theory, future data may validate them if B-modes are detected relatively soon.  Some of them, like $V\sim \phi$ and $V\sim \phi^{2/3}$ are shown in Fig.~\ref{CMBS42019}. The multi-field version of axion monodromy models \cite{Wenren:2014cga} may have smaller values of $n_s$, which would improve the agreement with the data. If these or other models are validated by the B-mode searches that are presently underway, such as BICEP2/Keck \cite{Ade:2018gkx}, this early detection of the primordial gravitational waves  will be a tremendous success. 

At present the error bars for the B-mode detection are too large to come to any conclusion in this respect, $\sigma(r) \sim 0.02$  \cite{Ade:2018gkx}. But during the next   5 - 10 years  it will become  $\sigma(r) \sim 0.005$, or even $\sigma(r) \sim 0.003$, depending on the level of delensing  that can be achieved in the future  \cite{Hui:2018cvg,Ade:2018gkx}.
Therefore the future missions will be needed to clarify any results of the current B-mode experiments.

\begin{figure}[t!]
\begin{center}
\hspace{-4mm}\includegraphics[scale=0.362]{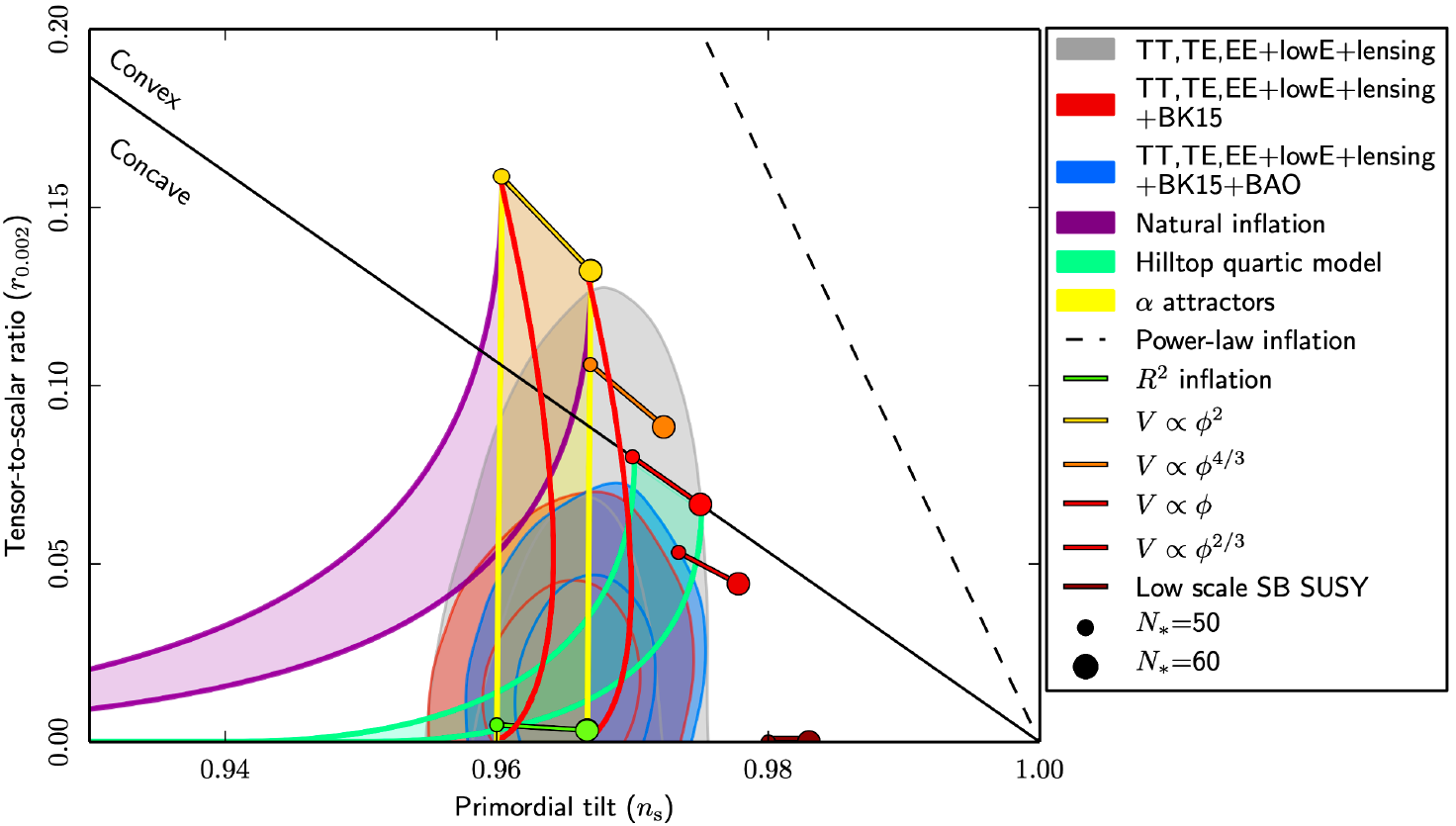}
\end{center}
\vskip -0.5cm 
\caption{\footnotesize  Yellow lines show predictions of the  T-models with $V\sim \tanh^{2}  {\varphi\over \sqrt {6\alpha}}$.   Red lines are for E-models, $V\sim \big (1- e^{ -\sqrt{{2\over 3 \alpha} }  \varphi}\big )^{2}$, for $N= 50$ and $N = 60$.  These two basic $\alpha$-attractor models together cover a significant part of the area favored by Planck 2018 and BICEP2/Keck (BK15).}
\label{F00}
\end{figure}

\begin{figure}[t!]
\begin{center}
\vskip 0.2cm
\includegraphics[scale=1.1]{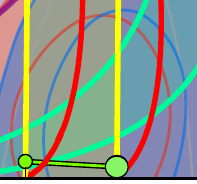}
\end{center}
\vskip -0.5cm 
\caption{\footnotesize This is the most relevant 
part of Fig.~\ref{F00}, which shows $\alpha$-attractor models with  $\alpha \lesssim 10$. Thick yellow lines correspond to the simplest T-models, thick red lines, slightly to the right,  correspond to the simplest E-models. A combination of these two models covers the main part of the central red (blue) $1\sigma$ ellipses favored by Planck 2018 and BK15. Red  ellipses show the results taking into account all available CMB-related data. This subset of the data was used in Planck 2018 for evaluation of inflationary models. Blue ellipses additionally take into account the data related to baryon acoustic oscillations.}
\label{Fshort}
\end{figure}

 Another set of simple models which are inside the Planck 2018 $2\sigma$ bounds on $n_s$ are $\alpha$-attractor models, see for example the gray stripe in Fig. \ref{CMBS42019}, which shows the prediction of the simplest T-model with potential  $V\sim \tanh^{2}  {\varphi\over \sqrt {6\alpha}}$ for   $47<  N < 57$, and also Fig.~\ref{F00} where the  red  lines show predictions of the simplest E-model with  $V\sim \big (1- e^{ -\sqrt{{2\over 3 \alpha} }  \varphi}\big )^{2}$,    and the  yellow lines correspond to  the T-model   $V\sim \tanh^{2}  {\varphi\over \sqrt {6\alpha}} $,  for $N= 50$ and $N = 60$.\footnote{ In this paper we use the Planck mass units $M_P=1$.} According to Planck 2018 \cite{Planck:2018jri}, the $\alpha$-attractor models with $\alpha \lesssim 10$ provide a good fit to the Planck data. We show these models in Fig.~\ref{F00} and Fig.~\ref{Fshort}.
  
However, one should be prepared to any outcome of B-mode experiments, especially if we have legitimate targets at $r \lesssim 10^{-2}$. This is the main subject of our investigation. 
A short summary of some of our results can be found in  \cite{Kallosh:2019eeu}.

Our  goal here is  to discuss the  simplest but physically motivated models, where a single parameter, or a combination of two parameters, is sufficient to fit all presently available data, and to identify some `future-safe' models, which have a fighting chance to describe and parametrize all data to be obtained in the next one or two decades. Note that a  comprehensive analysis of many inflationary models was performed in {\it Encyclop\ae dia Inflationaris} \cite{Martin:2013tda} and in the context of a CORE mission in  \cite{CORE:2016ymi}, based on Planck 2013 and 2015, respectively,  and an update of  {\it Encyclop\ae dia Inflationaris} based on Planck 2018  \cite{Planck:2018jri} is in preparation. A more recent analysis of single-field  inflationary models in \cite{Hardwick:2018zry} has emphasized the importance of  the 
decrease in the measurement uncertainty of the scalar spectral index.

To explain our motivation in a more detailed way we should note that it is possible to describe any set  of the 3  main parameters of inflationary perturbations, $A_{s}$,  $n_{s}$ and $r$, by tuning 3 parameters of a simple model $V = a\phi^{2} + b\phi^{3} + c\phi^{4}$ \cite{Destri:2007pv,Linde:2014nna}. Similarly, one can study a chaotic landscape of multifield potentials, and evaluate statistical probability of any outcome, without necessarily making sharp predictions, see e.g. \cite{Aazami:2005jf,Frazer:2011tg,Easther:2013rva,Bachlechner:2016mtp,Masoumi:2016eag,Linde:2016uec,Bjorkmo:2017nzd,Blanco-Pillado:2017nin}. One can also try to find multifield models predicting controllable amount of non-Gaussianity compatible with the Planck 2018 constraints, see e.g. \cite{Linde:1996gt,Lyth:2002my,Demozzi:2010aj,Linde:2012bt,Achucarro:2019mea,Bjorkmo:2019qno} and references therein.   These are legitimate possibilities.

However, it is difficult not to be excited and intrigued by the fact that dozens of models which were popular 10 years ago are already ruled out, whereas some of the models proposed more than 3 decades ago, such as the Starobinsky model \cite{Starobinsky:1980te}, the Higgs inflation model  \cite{Salopek:1988qh,Bezrukov:2007ep}, and the GL model  \cite{Goncharov:1985yu,Linde:2014hfa}, require just  a single parameter to successfully account for all presently available data. All of these models are shown in figure 2.2 from  PICO  \cite{Hanany:2019lle}. 

On the other hand,  a discovery of the gravitational waves in the range  $r \gtrsim 5 \times 10^{{-3}}$ would rule out all three of these models, whereas a non-discovery of  the gravitational waves in the range  $r \gtrsim  2 \times 10^{{-3}}$ would rule out the first two models in this list. How the predictions of  the Starobinsky model change if one adds to it some terms higher order in $R$? What will happen to the Higgs inflation if the potentials in these  theories have terms higher order in $\phi$? Many models of string theory inflation predict $r$ well below the level $5\cdot 10^{-4}$ to be reached by PICO. Do we have any other reasonable targets with $r \lesssim 10^{-2}$, or we just have optimistic expectations, similar to the expectations that the low energy supersymmetry will be discovered at LHC?

The starting point of our investigation are the results of Planck 2018 \cite{Planck:2018jri}. 
In Table~5 in \cite{Planck:2018jri}  there is a selection of  models which show the implications of  data for the most popular single-field slow-roll inflationary models, with a small number of free parameters.  Many of these models, such as the monomial models with $V \sim \phi^{2}$ or $\phi^{4}$,  are already  ruled out, 
but there are three classes of models which provide a very good fit to the Planck data. 

The first class of models includes the Starobinsky model, the Higgs inflation model, the GL model, and the large class of $\alpha$-attractors substantially generalizing all of these models \cite{Kallosh:2013yoa,Carrasco:2015uma,Kallosh:2017wnt}. We will describe these modes in detail in Sections \ref{alpha}, \ref{micro}. Predictions of $\alpha$-attractors, in the small $\alpha$ limit, are given by 
\be
1-n_s =  {2\over N}\, , \qquad r = {12\alpha\over N^2} \ . 
\ee
 Examples include  T-models with potentials $V\sim \tanh^{2n}  {\varphi\over \sqrt {6\alpha}} $ and  E-models with potentials  $V\sim \big (1- e^{ -\sqrt{{2\over 3 \alpha} }  \varphi}\big )^{2n} $.  The parameter $\alpha$ in these models  has a nice geometric interpretation in terms of the underlying hyperbolic geometry. 
 
 In general,  $\alpha$ may take arbitrary values. However, in section \ref{micro} we will discuss   7 especially interesting   discrete values  $3\alpha=7,6,5,4,3,2,1$, which are U-duality benchmarks associated with  M-theory, string theory, maximal ${\cal N}=8$  supergravity. They correspond to 7 different values of $ r $ in the range $10^{-3} \lesssim r \lesssim 10^{-2}$, which can be viewed as B-mode targets for the next round of CMB experiments.
 
 \begin{figure}[!h]
\vspace{-1mm}
\hspace{-3mm}
\begin{center}
 \includegraphics[scale=0.33]{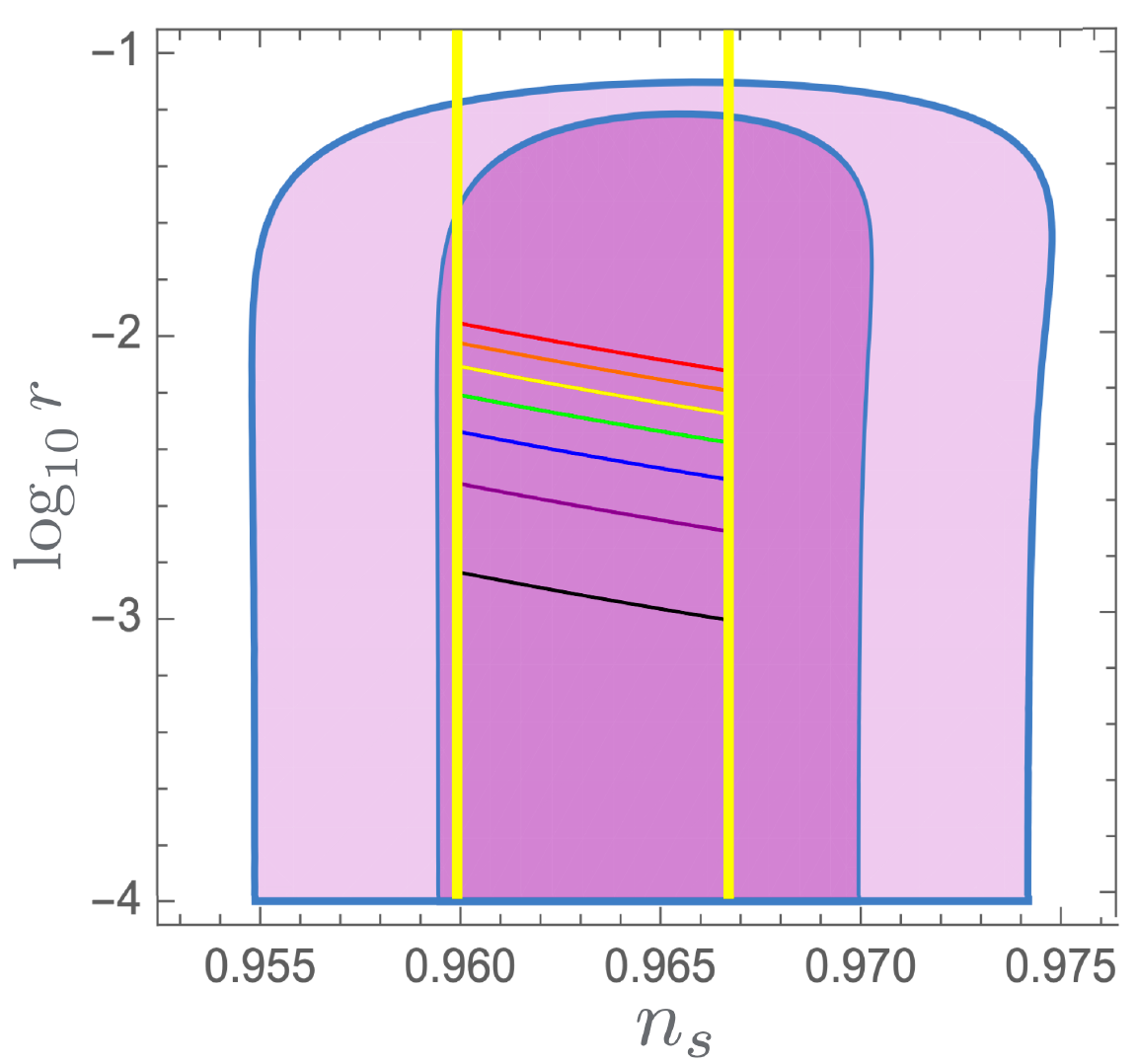}  \hskip 30pt
\includegraphics[scale=0.33]{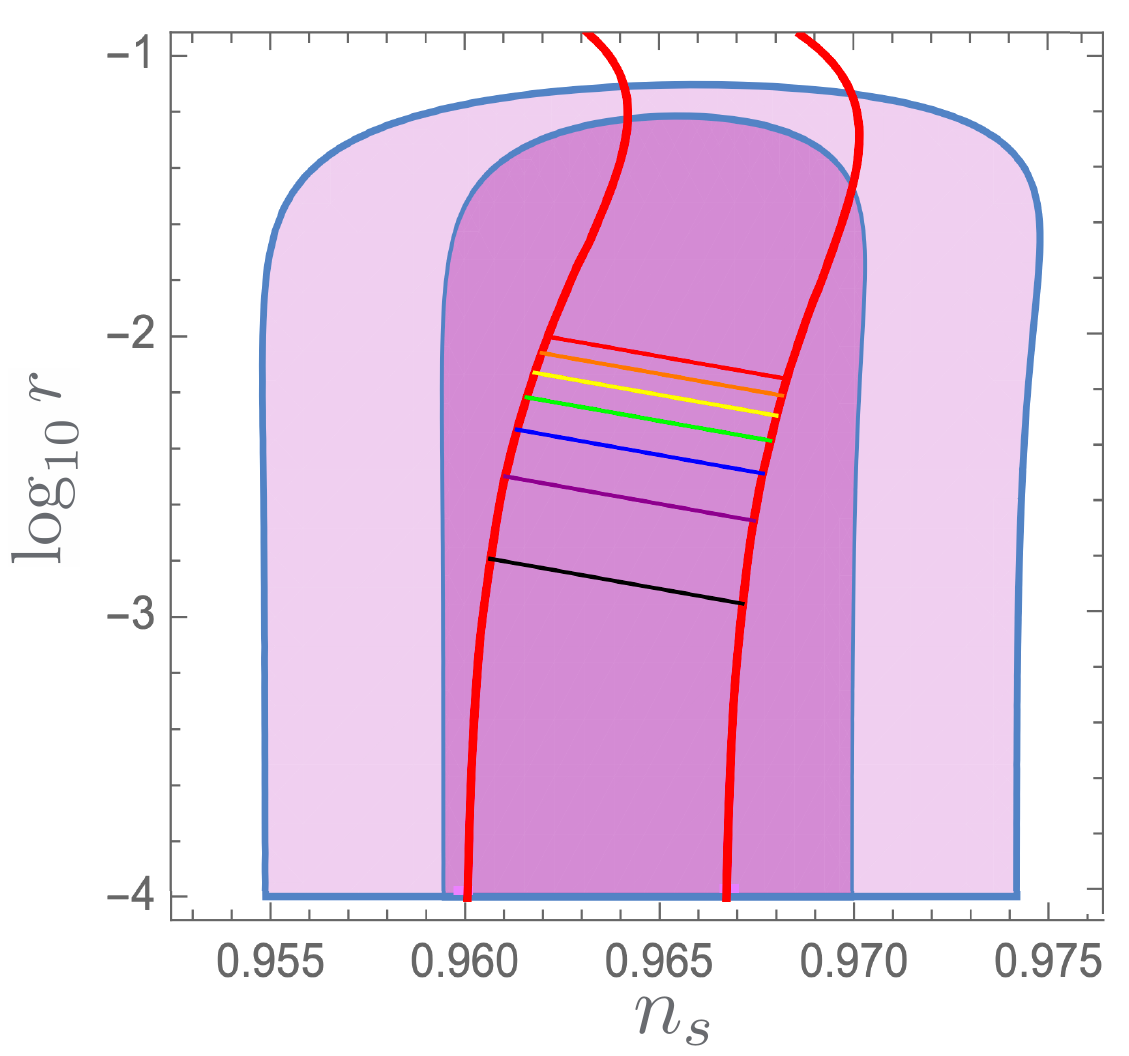} 
\end{center}
\vspace{-.12cm}
\caption{\footnotesize  U-duality benchmarks in  $\alpha$-attractor inflationary models originating from theories with maximal supersymmetry: M-theory, string theory, maximal supergravity. Simplest T-model is shown on the upper figure,  simplest E-models are shown on the lower figure. The 7-disk model \cite{Ferrara:2016fwe,Kallosh:2017ced} allows 7
 discrete values:   $3\alpha=7$ shown by a red line,  $3\alpha=6$ (orange), $3\alpha=5$ (yellow), $3\alpha=4$ (green), $3\alpha=3$  (blue), $3\alpha=2$ (purple) and     $3\alpha=1$ (black).  All other values of $\alpha$  originate from minimal supergravity models. Red ellipses show the Planck 2018 results taking into account  the CMB-related data including BK14. This subset of the data was used in Planck 2018 for evaluation of inflationary models.}
\label{7disk2}
\end{figure}

Some of these targets have other reasons to be examined. At $3\alpha = 6$ we would probe string theory fibre inflation \cite{Cicoli:2008gp,Kallosh:2017wku}, at $3\alpha =3$ we would probe the Starobinsky model~\cite{Starobinsky:1980te}, the Higgs inflationary model~\cite{Salopek:1988qh,Bezrukov:2007ep},  as well as the conformal inflation model~\cite{Kallosh:2013hoa}.  Finally, at $3\alpha =1$ we would probe the case of the maximal superconformal symmetry, as explained in Appendix \ref{conf}.  There is yet another target, at $\alpha = 1/9$, $r \sim 5\times 10^{{-4}}$, which corresponds to the GL model \cite{Goncharov:1985yu,Linde:2014hfa} shown by a purple dot in in figure 2.2 from  PICO  \cite{Hanany:2019lle}. This is a supergravity inflationary model  involving just a single superfield, which provided the first example of chaotic inflation with a plateau potential.

The second class of models favored by Planck 2018 includes the hilltop inflation models  with potentials $V\sim  1-{\varphi^{k}\over m^{k}} +... $ \cite{Linde:1981mu,Boubekeur:2005zm}. However,  the simplest models $V\sim   1-{\varphi^{k}\over m^{k}}  $ have the potential unbounded from below, and describe the universe collapsing immediately after inflation  \cite{Kallosh:2019jnl}. For $m \lesssim 1$, one can improve these models without modifying their inflationary predictions, but such models predict too low $n_{s}$ for  $k=2$ and $4$, so they are already ruled out. Meanwhile in the large $m$ limit all models $V\sim   1-{\varphi^{k}\over m^{k}}  $, for any $k$,  have  universal predictions for $n_{s}$ and $r$ coinciding with the predictions of the simple model with a linear potential $V \sim \phi$, as shown by the dark blue line at the right upper part of the green area in Fig.~\ref{CMBS42019}. According to \cite{Kallosh:2019jnl}, this universality, which could  be an attractive feature of hilltop inflation, is directly linked to the fundamental inconsistency of these models.

This does not mean that the full class of hilltop models is ruled out. However,   consistent generalizations of the  models $V\sim   1-{\varphi^{k}\over m^{k}}  $  for $m \gtrsim 10$ typically have very different predictions. 
One such model discussed in  \cite{Kallosh:2019jnl} is relatively well motivated (the Coleman-Weinberg model), but it does not seem to match the Planck data too well. Another model, with $V\sim  \bigl(1-{\varphi^{4}\over m^{4}}\bigr)^{2} $,  provides a better fit to the data, but it is not well motivated. Both of these models in the large $m$ limit predict much greater values of $r$ than the model $V\sim   1-{\varphi^{4}\over m^{4}}$.  Neither of them makes predictions reproducing the green area, which was supposed to describe hilltop inflation in the Planck, CMB-S4 and PICO figures.  We will not discuss these models here, and refer the readers to  \cite{Kallosh:2019jnl} for a detailed investigation of hilltop inflation after Planck 2018.

The third class of models favored by Planck 2018 includes Dp-brane inflation models with  $V\sim  1-{m^{k}\over \varphi^{k}}   + \dots   $   \cite {Martin:2013tda,Planck:2013jfk,Kallosh:2018zsi}, where $k = 7-p$, see section \ref{dbranes}.  Their simplest versions with  $V\sim  1-{m^{k}\over \varphi^{k}}$, which were called BI (brane inflation) in  \cite{Martin:2013tda},  are inconsistent for the same reason as the simplest hilltop models \cite{Kallosh:2019jnl}. Consistent generalizations of these models with potentials  $V\sim  (1+{m^{k}\over \varphi^{k}})^{{-1}} = {\varphi^{k} \over \varphi^{k}+m^{k}}$  were proposed in \cite{Kachru:2003sx} in the context of D3 brane inflation. These models  were generalized and   called KKLTI  (KKLT inflation)  in  \cite{Martin:2013tda}, and further developed in  \cite{Kallosh:2018zsi}.  
\begin{figure}[!h]
\vspace{-1mm}
\hspace{-3mm}
\begin{center}
 \includegraphics[scale=0.32]{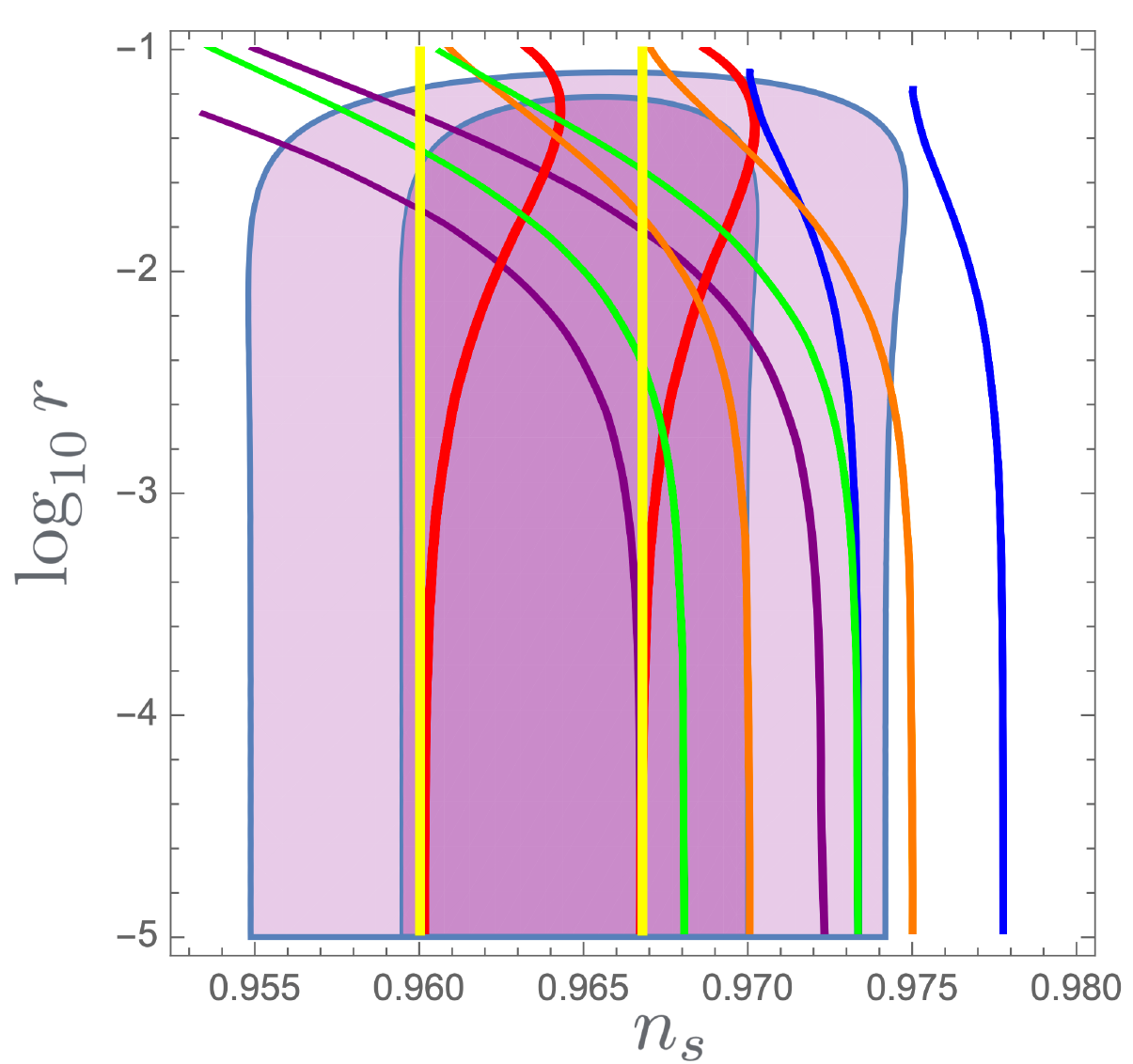}  \hskip 30pt
\includegraphics[scale=0.32]{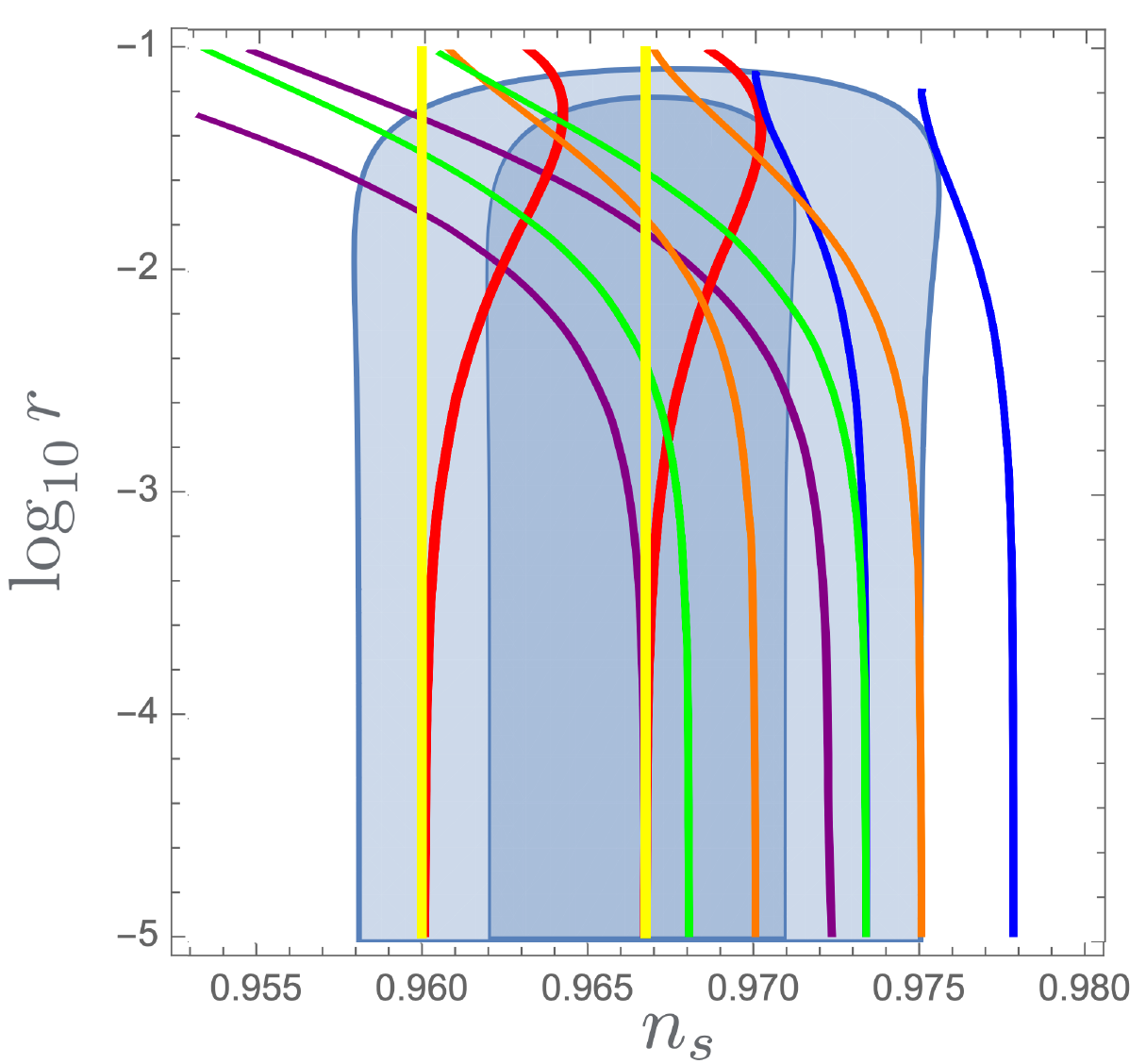} 
\end{center}
\vspace{-.12cm}
\caption{\footnotesize  A combined plot of the predictions of the simplest $\alpha$-attractor models and Dp-brane inflation  for $N = 50$ and $60$. From left to right, we show predictions of T-models, E-models,  ${\rm Dp}-{\overline {\rm Dp}}$  brane inflation with $p = 3, 4, 5, 6$. They are shown by yellow, red, purple, green, orange and blue lines correspondingly.  Red area shown in  the upper figure represents the Planck 2018 results taking into account  CMB-related data.  Blue area  shown in the lower figure additionally  takes into account the data related to BAO.}
\label{6pots}
\end{figure}

Predictions of $\alpha$-attractors and four  D-brane models with $p = 3$, 4, 5, 6 (i.e. with $k =4$,  3,  2, 1)\, can be represented by five vertical attractor stripes with $r\ll 1$ and
\be
 1-n_s =  {\beta\over N}\, , \qquad  \beta= 2, \  {5\over 3}, \   {8\over 5}, \  {4\over 3}, \  {3\over 2} \ . 
\label{simple}
\ee
As one can see from Fig. \ref{6pots}, they cover most of the $2\sigma$ area in the ($n_{s}$, $r$) space favored by Planck 2018. 
Moreover, to cover most of the $1\sigma$ area  favored by Planck 2018 it is sufficient to consider $\alpha$-attractors and   two  D-brane models with $p = 3$ and 5   \cite {Kallosh:2018zsi,Kallosh:2019jnl}.

Attractor $\beta$-stripes  \rf{simple} shown in Fig.~\ref{6pots}
 appear not only for $\alpha$-attractors and D-brane models, but also in a general pole inflation context introduced in \cite{Galante:2014ifa}, see also \cite{Terada:2016nqg} and 
sections  \ref{alpha}, \ref{general} of this paper. 
Pole inflation describes the cosmological attractors with the pole order $q$ in the kinetic term of the inflaton  field, see \rf{action}. In particular, $\alpha$-attractors are the pole inflaton models with $q = 2$, whereas D-brane inflation potentials (both KKLTI and BI) with  $k =4$,  3,  2,  1 belong to the class of the pole inflation potentials with $q = {5\over 3},\ {8\over 5}, \  {4\over 3}, \  {3\over 2}$ respectively. These models describe cosmological attractors which in the small $r$ limit predict $1-n_s =  {\beta\over N}$, where $\beta = {q \over q-1}$.

These results can be compared with the phenomenological parametrization of inflationary models based on an assumption that in ``natural'' models of inflation one may expect $1-n_{s}= {p+1\over N}$, where $p$ is some phenomenological parameter   \cite{Mukhanov:2013tua,Creminelli:2014nqa, Abazajian:2016yjj}.  In  our paper, we use   $\beta$ instead of $p+1$ to avoid confusion with  $p = 3, 4, 5, 6$ in Dp-brane inflation, where the use of the letter $p$ in  Dp is a long accepted standard. 

As we will see,  pole inflation provides a convenient theoretical framework for the phenomenological parametrization used  in  \cite{Mukhanov:2013tua,Creminelli:2014nqa, Abazajian:2016yjj}. In particular, we will show that the characteristic scale of inflation introduced in  \cite{Abazajian:2016yjj} is directly related to the  residue $a_{q}$ at the pole of the inflaton kinetic term,  see section \ref{Char}.  On the other hand, our results obtained in section \ref{general} show that we may not need to have a large  continuous range of parameters $\beta$: the predictions of the cosmological attractors described by the two stripes $\beta = 2$ and $\beta =5/3$ completely cover the $1\sigma$ region   in the ($n_{s}$,~$r$) space favored by Planck 2018, see Fig.~\ref{QQ2}.

While we are unaware of any  specific targets for $r$ in D-brane inflation models and general pole inflation models with $q \not = 2$, the search of the B-modes, in combination with the improvement of the precision in  the measurements on $n_{s}$, may be very important to distinguish different versions of these models from  $\alpha$-attractors and to  get a better understanding of the post-inflationary evolution of the universe, including reheating,  affecting the required value of the e-foldings $N$ in all of these models.\footnote{The standard assumption is that  $N$ can be in the range from 50 to 60 (or from 47 to 57), but  this range can be more broad, depending on the mechanism of reheating. For example, for  quintessential $\alpha$-attractors with gravitational reheating, the required value of the e-foldings  $N$ can be greater than  in more conventional models by $\Delta N \sim 10$, which increases the predicted value of $n_{s}$ by about $0.006$  \cite{Akrami:2017cir}. This additional  increase can be greater than the Planck $1\sigma$ error bar  for $n_{s}$.}

\section {\boldmath{Inflationary $\alpha$-attractor models}}\label{alpha}
We would like to explain here that in general class of $\alpha$-attractor models  the information about observables $n_s$ and $r$ is codified in their kinetic terms, under specific conditions. For example, the models have to be in their attractor regime, etc. The reason why $\alpha$-attractors  have specific benchmarks, to be discussed later,  is this fact that the observational data are defined by  kinetic terms of the theory. Kinetic terms for scalars   are often defined by the symmetries of the theory, which may be  broken by the potential. For example, the kinetic terms of scalars in maximal $\cN=8$ supergravity is defined by U-duality symmetry, \E.

 It is convenient to explain this feature using the `pole inflation' version of $\alpha$-attractors~\cite{Galante:2014ifa}.

\subsection{\boldmath $\alpha$-attractors and pole inflation: E-models}
There are many different ways to introduce $\alpha$-attractors. In the context of this paper, it is useful to start with  the pole interpretation of these models \cite{Galante:2014ifa}
\be
{\cal L} = {\cal L}_{\rm kin} - V  = - {1\over 2} {a_q\over \rho^{q}} (\partial \rho)^2- V(\rho) \ .
\label{action}\ee
Here the pole of order $q$ is at $\rho=0$ and  the residue at the pole is $a_q$. If the potential is regular near the pole,
\be 
V = V_0 ( 1- c \rho + \ldots) , \qquad c > 0 \ ,
\label{V}\ee
one finds that inflation occurs in a small vicinity of the pole. Inflationary predictions $n_s$ and $r$ depend on $q$, on $a_q$, on the number of e-foldings $N$, and, in general, on the constant $c$ in the  potential. 
 
As an example, let us first consider the simplest and the most important case $q=2$, with $a_{2} \equiv {2\over 3\alpha}$.  In that case one can make a change of variables $\rho = e^{-\sqrt {2\over 3\alpha}\varphi}$. The theory \rf{action} after the transformation represents a canonical field $\varphi$ with action
\be
{\cal L} = {\cal L}_{\rm kin} - V  = - {1\over 2}  (\partial \varphi)^2- V(e^{-\sqrt {2\over 3\alpha}\varphi}). 
\ee
We called these models E-models, because of the exponential change of variables $\rho = e^{-\sqrt {2\over 3\alpha}\varphi}$. Inflation occurs at large positive values of the canonically normalized field $\varphi$, where the potential  is given by
\be 
V = V_0 \bigl( 1 -  c \, e^{-\sqrt {2\over 3\alpha}\varphi}  + \ldots \bigr) \ .
\label{Vcan}\ee
It  approaches the plateau from below. The canonical kinetic term $- {1\over 2}  (\partial \varphi)^2$ is invariant under the constant shift of the inflaton, and the constant $c$ can be absorbed into a redefinition of the exponential term.  Therefore the theory at $\sqrt {2\over 3\alpha}\varphi \gg 1$ is equivalent to the one with a potential
\be V = V_0 \bigl( 1 - e^{-\sqrt {2\over 3\alpha}\varphi}  + \ldots \bigr) \ .
\label{Vcan1}\ee
But this is not a good potential because it is unbounded from below at $\varphi \to -\infty$. 
The simplest example of a consistent inflationary potential  in this context is provided by $V =V_{0}(1-\rho)^{2}$. In the canonical variables it is given by 
\be\label{E}
 V = V_0 \Bigl(1 - e^{-\sqrt {2\over 3\alpha}\varphi}\Bigr)^{2} . 
\ee
For $\alpha = 1$  this potential coincides with the potential of the Starobinsky model. The main difference is that the action of the original Starobinsky model by design represents the Einstein  action with an additional term $R^{2}$, with a very large coefficient in front of it. But if one is allowed to add the large term $\sim R^{2}$, one may also consider general terms $F(R)$, which may change the structure of the potential. The situation is similar to what happens in the theory of a scalar field $m^{2}\phi^{2}/2$ if one  replaces it by an arbitrary potential $V(\phi)$: Inflation remains possible for an appropriate choice of $V(\phi)$, but inflationary predictions depend on the choice of the potential. This is related to the so-called $\eta$ problem.

Meanwhile in the context of $\alpha$-attractors, the asymptotic expression for any potential $V(\rho)$ growing but remaining non-singular at $\rho \to 0$ continues to be given by equation \rf{Vcan1}. This explain stability of the predictions of $\alpha$-attractors with respect to considerable modifications of $V(\rho)$, including possible quantum corrections \cite{Kallosh:2016gqp}. 

Some part of this stability is a general property of the theories \rf{action}, but the possibility to absorb the constant $c$ in \rf{V} into a shift of the field $\varphi$ is a unique property of the models with $q = 2$. In this case the residue of the pole, introduced in \cite{Galante:2014ifa},  $a_2= {2\over 3\alpha}= {1\over |\mathcal{R}_K|} $ has a geometric origin. It was explained in  \cite{Ferrara:2013rsa,Kallosh:2015zsa} that the \K\, curvature of the underlying moduli space is $\mathcal{R}_K= -{2\over 3\alpha}$. 

One can also absorb the constant $c$ in the potential into $\rho$ for an arbitrary $q$
\be\label{redef0}
\tilde \rho\equiv  c\rho \ .
\ee
In such case
\bea
{\cal L}   &=& - {1\over 2} a_q  {(\partial \rho)^2\over \rho^{q}} -  V_0 ( 1- c \rho + \ldots) \\ \nonumber
&=& - {1\over 2}  c^{q-2} a_q { (\partial \tilde \rho)^2\over \tilde \rho^{q}} -  V_0 ( 1-  \tilde \rho + \ldots)
 \ .
\label{action1}\eea 
For  $q \not = 2$ removing $c$ from the potential results in the rescaling of the residue of the pole
\be
\tilde a_q = c^{q-2} a_q \ .
\label{redef}\ee
Thus, we could have started with a potential with $c=1$ and a redefined residue of the pole, as shown in eq. \rf{redef}
\be
{\cal L}   =  - {1\over 2}  \tilde a_q { (\partial \tilde \rho)^2\over \tilde \rho^{q}} -  V_0 ( 1-  \tilde \rho + \ldots)
 \ .
\label{action2}\ee 
Note that only in $q=2$ case where we have the hyperbolic geometry, the residue of the pole $a_2= {2\over 3\alpha}= {1\over |\mathcal{R}_K|} = \tilde a_2$ has a geometric meaning, and we see that removing the constant $c$ from the potential does not change the residue. In all other cases the original value $a_q$ or the rescaled one $\tilde a_q$ are not associated with any geometry and can be used for the purpose of a convenient description of the inflationary predictions of these models.

Explicit expressions for the spectral index $ n_s$, the tensor-to-scalar ratio $r$, and the amplitude of perturbations $A_{s}$ in leading order in $1/N$ at small $\alpha$ were 
  derived in  \cite{Galante:2014ifa} for $q\neq 1$. We use the following notation here for the order of the pole $q$ in eq. \rf{action}
 \be
 q= {\beta\over \beta-1}\, ,  \qquad \beta= {q\over q-1} \ ,
  \ee
 and we find  
  \be
   n_s = 1 -  \frac\beta N \,, \quad
  r= 8 \,  \tilde a_q^{\beta-1} \Big ({{\beta-1}\over {N} }\Big )^{\beta}\, , \quad A_{s}  = {2 V_{0}\over 3 \pi^{2}r} \ .
  \label{pole}
\ee
Thus, at  $q=2$, all dependence on the parameter in the potential $c$ in \rf{action} in $r$ 
disappears without the need to redefine the residue of the pole, it is by preserving the one defined by geometry!
For  $q=\beta =2$,  $a_{2} \equiv {2\over 3\alpha}$, these predictions are
\be
   n_s = 1 - \frac{2}{N}   \,, \quad
  r=\frac{12\alpha}{N^{2}}  \, , \quad A_{s}  = V_{\alpha} \, { N^{2}\over 18\pi^{2}} \ .
  \label{pole2}
\ee
where $V_{\alpha} = {V_{0}\over  \alpha}$. This means that for sufficiently small $\alpha$ and large $N$ all  $\beta=q=2$ $\alpha$-attractor models have the same values of $n_s$ and a value of $r$ which is independent on the potential, and depends only on $\alpha$. We will explain in Section \ref{micro} that these $\beta=2$ $\alpha$-attractor models originate from the hyperbolic geometry, with the curvature   $ \mathcal{R}= - {2\over 3\alpha}$  \cite{Ferrara:2013rsa}.

As long as the prediction $n_s = 1 - \frac{2}{N}$ provides a good fit to the Planck 2018 data, the single parameter that we need to adjust is  $V_{\alpha}  \sim 10^{{-10}}$, which provides the amplitude of perturbations  consistent with Planck normalization. And then, by adjusting $\alpha$ we can describe any value of $r$ found by the B-mode searches.

Meanwhile the situation with $q\not =2$ is slightly more complicated. The values  of $r$  and  $A_{s}$  for $q \not = 2$ depend on $c$, i.e. on the functional form of the potential, see \rf{redef0}, \rf{redef}, \rf{pole}, even though this dependence can be absorbed into the field redefinition. For a more detailed discussion of related issues see \cite{Galante:2014ifa,Terada:2016nqg} and 
section \ref{general}.

\subsection{T-models}

From this perspective it may be important that the original
$\alpha$-attractor models derivable from supergravity always have the pole of order $q=2$,  due to hyperbolic geometry \cite{Kallosh:2013yoa,Kallosh:2015zsa,Carrasco:2015uma,Ferrara:2016fwe,Kallosh:2017ced,Kallosh:2017wnt}.
\be\boxed{
\rm{ hyperbolic\, \,  geometry} \qquad \Rightarrow \qquad q=2 }
\ee
Prior to discussing it, it is important to introduce yet another class of $\alpha$-attractors, T-models. The simplest example is given by the theory
 \be
 {1\over \sqrt{-g}} \mathcal{L} = { R\over 2}   -  {(\partial_{\mu} \phi)^2\over 2\bigl(1-{\phi^{2}\over 6\alpha}\bigr)^{2}} - V(\phi)   \,  .
\label{cosmoA}\ee
Here $\phi(x)$ is the scalar field, the inflaton.  Once again, the kinetic term is  singular, but now the singularity is at $|\phi| = \sqrt{6\alpha} $. Instead of the variable $\phi$, one can use a canonically normalized field $\vp$ by solving the equation ${\partial \phi\over 1-{\phi^{2}\over 6\alpha}} = \partial\vp$, which yields $
\phi = \sqrt {6 \alpha}\, \tanh{\varphi\over\sqrt {6 \alpha}}$.
The full theory, in terms of the canonical variables, becomes
 \be
 {1\over \sqrt{-g}} \mathcal{L} = { R\over 2}   -  {(\partial_{\mu}\varphi)^{2} \over 2}  - V\big(\sqrt {6 \alpha}\, \tanh{\varphi\over\sqrt {6 \alpha}}\big)   \,  .
\label{cosmoqq}\ee
 Asymptotic value of the potential at large $\varphi>0$ is given by
\be\label{plateau}
V(\vp) = V_{0} - 2  \sqrt{6\alpha}\,V'_{0} \ e^{-\sqrt{2\over 3\alpha} \varphi } \ .
\ee
Here $V_0 = V(\phi)|_{\phi =  \sqrt {6 \alpha}}$ is the height of the plateau potential, and $V'_{0} = \partial_{\phi}V |_{\phi = \sqrt {6 \alpha}}$. As in the E-models, the coefficient $2  \sqrt{6\alpha}\,V'_{0}$ in front of the exponent  can be absorbed into a redefinition (shift) of the field $\varphi$. Therefore all inflationary predictions of this theory in the regime with $e^{-\sqrt{2\over 3\alpha} \varphi } \ll 1$ are determined only by two parameters, $V_{0}$ and $\alpha$.

The amplitude of inflationary perturbations $A_{s}$ in these models matches the Planck normalization for  $ {V_{0}\over  \alpha} \sim 10^{{-10}}$.  For the simplest  model $V = {m^{2}\over 2} \phi^{2}$, belonging to a class of T-models  with the potential symmetric with respect to $\phi \to -\phi$, one finds
\be\label{T}
V =  3m^{2 }\alpha \tanh^{2}{\varphi\over\sqrt {6 \alpha}} \ .
\ee
Then  the condition $ {V_{0}\over  \alpha} \sim 10^{{-10}}$ reads $ m  \sim   0.6 \times10^{{-5}}$. 

We should note that even though the predictions for large $N$ and small $\alpha$ are rather well defined, the value of $N$ itself does depend on the mechanism of reheating and post-inflationary equation of state, which is reflected in the uncertainty of the choice between $N\sim 50$ and $N \sim 60$. Also, predictions of different versions of $\alpha$-attractors converge to their target \rf{pole2} in a slightly different way.  In Fig. \ref{F00} one can see that  the predictions of the simplest T-models and E-models coincide in the two opposite limits, $\alpha \to 0$ and at $\alpha \to \infty$. Meanwhile for intermediate values of $\alpha$ the E-models predict somewhat higher values of $n_{s}$. As a result, a combination of these two models cover a significant part of the range of  $n_{s}$ and $r$ favored by Planck 2018, see Figs.~\ref{F00} and \ref{Fshort}. 

 \begin{figure}[t!]
\begin{center}
 \hspace*{-4mm} \includegraphics[width=9.3cm]{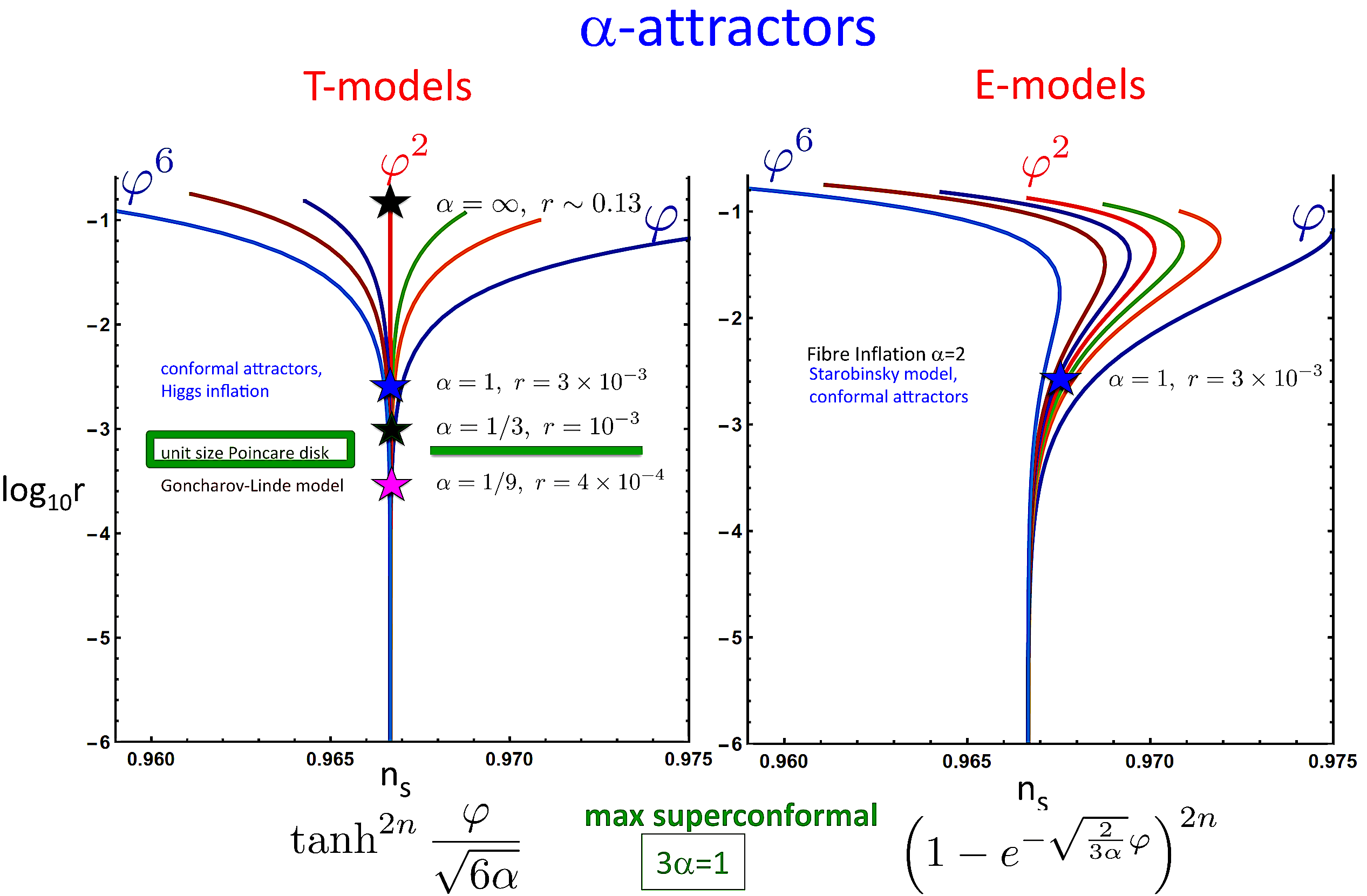}
\end{center}
 \vspace*{-4mm}
\caption{\footnotesize Values of $n_s$ and $r$ for simplest  T-models \rf{T} and E-models \rf{E}. Here $n =  {1\over 2}, {3\over 4}, {7\over 8}, 1, {3\over 2}, 2, 3$, starting from the right, increasing to the left, with the purple line for $n=1$ in the middle, for $N=60$ \cite{Carrasco:2015pla}. The predictions of these models  interpolate between the predictions of various polynomial models $\vp^{2n}$  at very large $\alpha$ and the vertical attractor line $n_{s}=1-2/N$ for $\alpha\leq 1$. Note that E-models tend to have slightly higher values of $n_{s}$ than  T-models at $r>10^{-3}$. 
}
\label{alpha1}
\vspace{-0.3cm}
\end{figure}

As we already mentioned,  predictions of the models  E-models \rf{E} for $\alpha = 1$  coincide with the predictions of the Starobinsky model. Similarly, the predictions of the T-model with the potential $V \sim \tanh^{4}{\varphi\over\sqrt {6 \alpha}}$  for $\alpha = 1$ nearly coincide with the predictions of  the   Higgs inflation. However, unlike Higgs inflation, predictions of  $\alpha$-attractors at small are rather stable with respect to the change of the potential $V(\phi)$ and allow much greater flexibility with respect to the tensor to scalar ratio $r$ by changing $\alpha$.  In this respect, $\alpha$-attractors are more ``future-safe'', allowing to   describe and parametrize various outcomes of the B-mode searches.

On the other hand,  predictions of $\alpha$-attractors at greater values of $\alpha$ and $r$ allow some variability,  see the behavior of  $n_s$ and $r$ at all $\alpha$ for $N=60$ in Fig.~\ref{alpha1}, taken from \cite{Carrasco:2015pla}.   At $\alpha \gtrsim 1$ they are not given by the attractor equation \rf{pole2}, they approach the predictions of the potential $V \sim \varphi^{2n}$ in the limit $\alpha \rightarrow \infty$. If inflationary B-modes are detected above $r\sim 10^{-3}$ and precision in $n_s$ improves, one may use these choices of $\alpha$-attractor models to find the best one.

\begin{figure}[!h]
 \vspace*{3mm}
\begin{center}
\hskip 0.3cm\includegraphics[width=5cm]{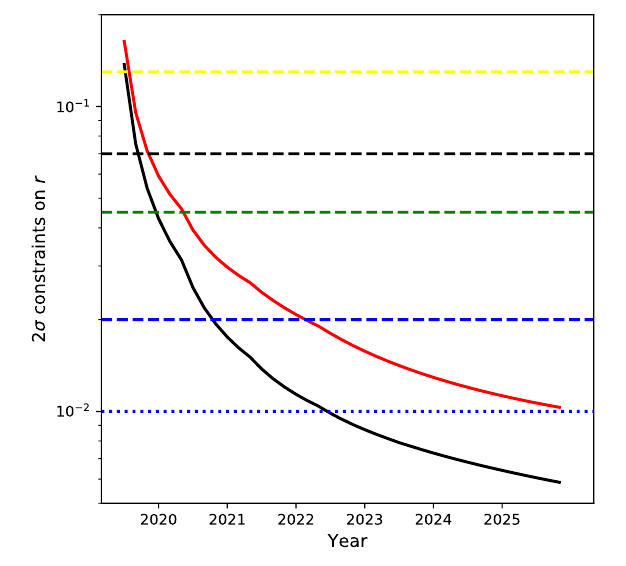}
\caption{\footnotesize The black and red curve represents the AliCPT $2\sigma$  limits on $r$ where in the simulations they have considered  residual foreground of 1\% (black) and 10\% (red). The black dashed line is the  limit from BICEP/Keck Array and Planck collaborations, 2016. 
The yellow dashed line shows the already excluded $\varphi^2$ model, the green dashed line shows the predictions from the axion monodromy model with potential  $\varphi^{3/2}$ \cite{McAllister:2014mpa}. The blue dashed and dotted lines are for the simplest $\alpha$-attractor model $\tanh^2  {\varphi\over \sqrt {6\alpha}}$   \cite{Kallosh:2013yoa} with $\alpha=7$, $\alpha=3$, respectively. The number of e-folds is taken to be $N = 60$.
 }
\label{Ali}
\end{center}
\vspace{-0.3cm}
\end{figure}

One of the important features of  $\alpha$-attractors is the fact that they depend on one parameter $\alpha$, and therefore this parameter  can be, eventually,  inferred from the  observational data, or better bounds on it can be obtained.
Consider, for example,  the  description of the Tibet's experiment probing primordial gravitational waves, see Fig. \ref{Ali} here, in  \cite{Li:2017drr,Li:2018rwc}.  They show the  
scheduled AliCPT sensitivity of the measurements on $r$ superposing it with the  theoretical predictions of three simple inflationary models. 

Note that the simple targets here, below the predictions of the most studied string theory axion monodromy model \cite{Silverstein:2008sg,McAllister:2008hb,McAllister:2014mpa} shown by the dashed green line, are given for T-models with for $r\approx 2.3\cdot 10^{-2}$ with $\alpha=7$ and $r\approx  10^{-2}$ with $\alpha=3$.

 \subsection{Geometry of ${\alpha }$-attractors} \label{proposal}
 
 Geometric features of inflationary ${\alpha }$-attractors originate from  supergravity \cite{Ferrara:2013rsa,Kallosh:2015zsa}. In case of $\cN=1$ supergravity we can start with the \K\, potential in the form $K= -3\alpha \ln (1-Z\bar Z)$, $Z\bar Z<1$,  and the metric 
 in the line element $ds^2= g_{Z\bar Z} dZ d\bar Z$ is \be
 g_{Z\bar Z} = \partial_Z \partial_{\bar Z} K(Z, \bar Z)= {3\alpha \over (1-Z\bar Z)^2} \ .
 \label{metric}\ee
  The meaning of `scalars are coordinates of the moduli space' is the following: we identify the kinetic terms of the complex scalar $Z$ from the moduli space metric \rf{metric}:
\be
- {\cal L}_{kin} = g_{Z\bar Z} \partial Z \partial \bar Z=  3\alpha {\partial Z \partial \bar Z\over (1-Z\bar Z)^2} ,  \quad g_{Z\bar Z}=  {3\alpha \over (1-Z\bar Z)^2}   ,
\label{kinZ} \ee
 and the part of the action,  gravity + kinetic term for scalars, in units $M_P=1$,   is
\be
{\cal L} = {1\over 2}R - {3\alpha \over (1-Z\bar Z)^2} \partial Z \partial\bar Z  \ .
\label{N1}\ee
Once we have a geometry and a metric, we can define the curvature
  which in our case is a negative constant
 \be
 \mathcal{R}_{{K}}= g^{Z\bar Z} \partial_Z \partial _{\bar Z} ( \log g_{Z\bar Z})= -{2\over 3\alpha} \ .
\label{curv} \ee
Since the corresponding geometry is a \K\, geometry, $g_{Z\bar Z} = \partial_Z \partial_{\bar Z} K(Z, \bar Z)$,  the curvature $\mathcal{R}_K= -{2\over 3\alpha}$ is known as a \K\, geometry curvature. 
\begin{figure}[ht!]
\begin{center}
\includegraphics[width=5cm]{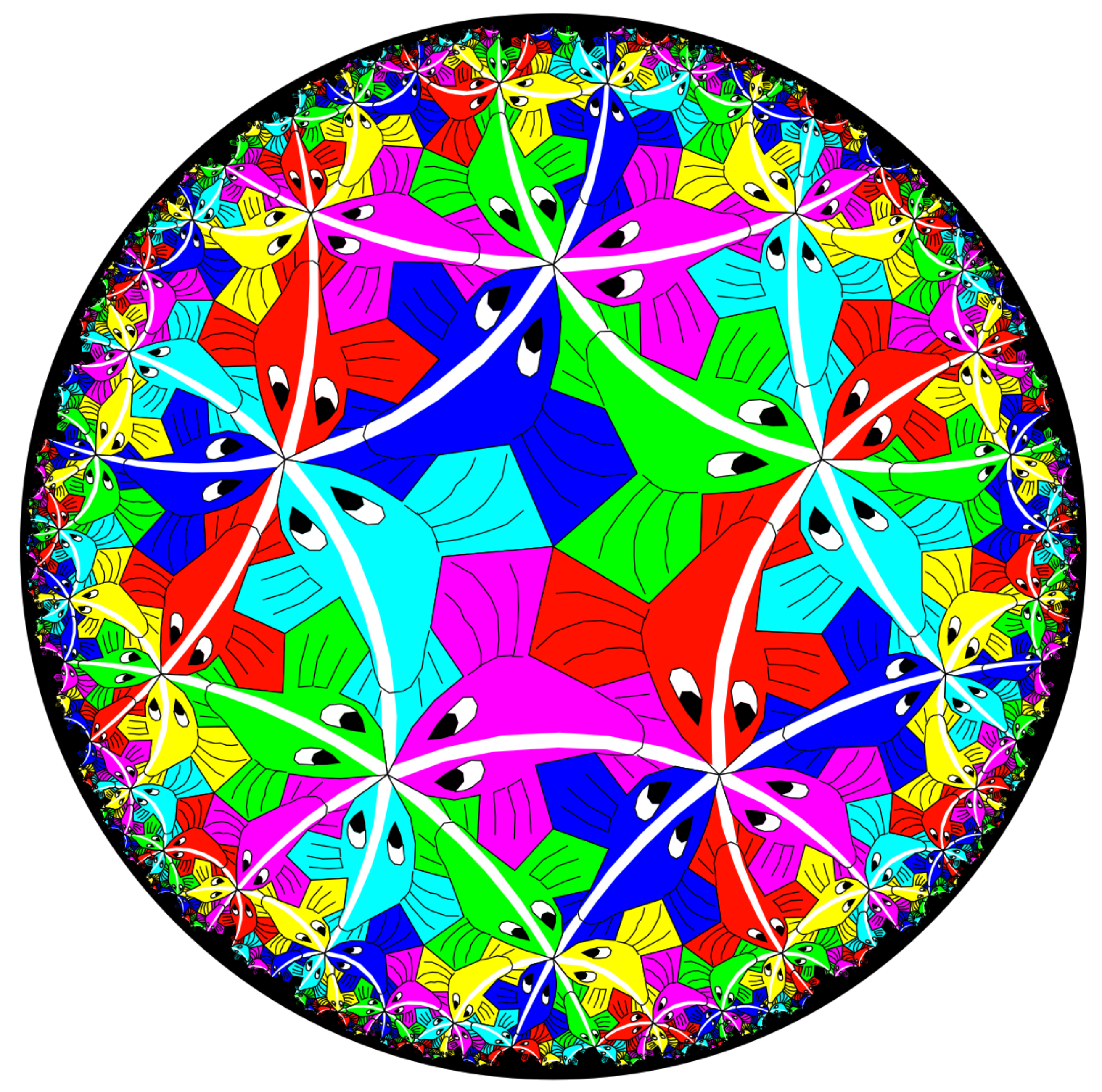}
\caption{\footnotesize 
A computer generated picture by D. Dunham inspired by
Escher's picture Circle Limit III  presents a Poincar\'e disk model of a hyperbolic geometry. The M\"obius symmetry of the geometry is illustrated here via a configuration of fishes.} 
\label{f4}
\end{center}
\vspace{-0.3cm}
\end{figure}

We now switch from Cartesian coordinates $Z= x+iy$ to polar coordinates on the disk, with some rescaling 
\be
 x +i y ={1\over  \sqrt {3\alpha}}  r e^{i\theta} \ .
 \ee  
The moduli space metric becomes
\be\label{metr}
ds^2=  g_{Z\bar Z} d Z d \bar Z=  {d r^2 +r^2 d\theta^2\over (1-{r^{2}\over 3\alpha})^{2}} \, , \qquad r^2< 3\alpha
\ee
where the original geometric constraint $Z\bar Z<1$ becomes $r^2< 3\alpha$.  Here $r$ has  a clear interpretation of the radial coordinate of the disk, whereas $\theta$ is an angular coordinate of the disk. The boundary cycle 
\be
r^2= 3\alpha
\ee
is not part of the disk model of the hyperbolic plane, it is called {\it absolute}.

The Poincar\'e disk is often depicted in Escher pictures, see Fig.~\ref{f4} here and \cite{Kallosh:2015zsa} for more details on this.
The radius of the Poincar\'e disk, which we  call  $R_{\rm Escher}$,
is defined as
\be
 R_{\rm Escher}^2 \equiv 3\alpha\, ,  
\ee
since the radial coordinate $r$ is constrained by  
\be
r^2<  R_{\rm Escher}^2 \ .
\ee
 $R_{\rm Escher}$ is related to the \K\, curvature as 
 \be
 \mathcal{R}_{K}=  -{2\over R^{2}_{\rm Escher}} \ .
\ee
 In particular for $3\alpha =1$ we recover the unit size Poincar\'e disk with $R^2_{\rm Escher} =3\alpha =1$ and metric \rf{metr}:
 \be 
ds^2|_{3\alpha =1}=  {d x^2+d y^2\over (1-{ x^2+ y^2} )^2}= {d r^2 +r^2 d\theta^2\over (1-r^{2})^{2}}  \ .
  \label{Escher}\ee
The scalar kinetic term in \rf{kinZ} in polar variables becomes
\be
- {\cal L}_{kin} = {(\partial r)^2 +r^2 (\partial \theta)^2\over (1-{r^{2}\over 3\alpha})^{2}}\  .
\label{polarL} \ee
At $\theta$=0, which corresponds to a stabilization of the angular variable during inflation, 
this is a slice of the Escher's picture, at fixed angular direction. It is useful to compare this kinetic term for the scalars with the expression in \rf{cosmoA} where $r\equiv {\phi\over \sqrt 2}$ and in \rf{cosmoqq} where $
\phi = \sqrt {6 \alpha}\, \tanh{\varphi\over\sqrt {6 \alpha}}$, so that $\varphi$ is a standard canonical field of a single-field inflationary model.

\section{\boldmath {From the microscopic theory of ${\alpha }$-attractors to B-mode targets}}\label{micro}

\subsection{\boldmath {$\cN=1$ d=4 supergravity predictions for observables in $\alpha$-attractors}}
Supergravity moduli space must be  described by the  \K\, geometry, it is a necessary condition for supersymmetry. A class of $\cN=1$ supergravities beyond the  general class of $\alpha$-attractors  is based on a \K\, potential
\be
K= - 3\alpha \ln (1-Z\bar Z)\, ,  \qquad \alpha > 0 \ .
\ee
The corresponding metric defining the kinetic term for scalars is a second derivative of the  \K\, potential
$
 g_{Z\bar Z} \equiv  \partial_Z \partial_{\bar Z} K   =  { 3\alpha\over (1-Z\bar Z)^2} $. Thus microscopic inflationary models of $\alpha$-attractors are based on {\it hyperbolic geometry}. This means that the theory has a kinetic term for complex scalars, which are coordinates of the Poincar\'e disk, $Z\bar Z<1$, the kinetic term is of the form shown in eq. \rf{kinZ}.
And as explained above,  we can define the \K\, curvature
 as $\mathcal{R}= g^{Z\bar Z} \partial_Z \partial _{\bar Z} ( \log g_{Z\bar Z})$ which in our case is a negative constant, $\mathcal{R}= -{2\over 3\alpha}$. The scalars in these models are coordinates of the coset space $SU(1,1)\over U(1)$. For related supergravity models of inflation in no-scale supergravity see \cite{Ellis:2019bmm} and references therein.

 The potential is usually a function of the disk coordinate $Z$ and is chosen so that
 the inflationary trajectory is stabilized at $Z=\bar Z$, $\theta=0$,  and  the real part of the complex scalar $Z$ is an inflaton.
\be
Z= \bar Z= {\phi\over \sqrt{3\alpha}} = \tanh {\varphi\over \sqrt {6\alpha}} \ ,
\ee
and the kinetic term $-3\alpha {\partial Z \partial \bar Z\over (1-Z\bar Z)^2}\Big| _{Z=\bar Z}$ becomes
\be
- {1\over 2} {(\partial \phi)^2\over (1-{\phi^2\over 6\alpha})}= 
 -{1\over 2} (\partial\varphi)^2 \ .
\ee
Microscopically we have a clearly identified fundamental parameter: a negative curvature of the hyperbolic moduli space,
$
\mathcal{R}= - {2\over 3\alpha}
$.
T-models have a potential $V_T\sim (Z\bar Z)^n\sim \tanh^{2n} {\varphi\over \sqrt {6\alpha}}$.

E-models are simpler when the half
-plane variables $T= {1+Z\over 1-Z}$   are used.  The moduli space metric  in this case and the kinetic term are  
\be
ds^2= 3\alpha {dT d\bar T \over (T+\bar T)^2}= g_{T\bar T} dT d\bar T , \quad  {\cal L}_{\rm kin} =- 3\alpha {\partial T \partial \bar T \over (T+\bar T)^2} \ , \label{half}
\ee
and the potential is
\be
V_E\sim (1-T )^{2n}\sim \left(1-e^{- \sqrt{ 3\alpha\over 2} \varphi}\right)^{2n} . 
\ee

The  kinetic terms in disk or half-plane variables with {\it an arbitrary $\alpha$}  have a simple embedding into minimal $\cN=1$ d=4 supergravity as shown in  \cite{Kallosh:2013yoa,Carrasco:2015uma,Kallosh:2017wnt}. Therefore these models compatible with the data are also compatible with $\cN=1$ supergravity, so in this context minimal   $\cN=1$   supergravity implies 
\be  
 n_s \approx 1-{2\over N}\, ,  \qquad r \approx {12\, \alpha \over N^2} \ ,    
\ee
where $\alpha$ can take arbitrary values.
In particular,  $\cN=1$ supergravity  is compatible with any value of  $ r \lesssim 7\cdot 10^{-2}$, which is the current experimental bound on $r$.

\subsection{U-duality benchmarks }

A U-duality symmetry is a fundamental symmetry in M-theory, string theory, maximal d=4 $\cN=8$ supergravity. In d=4 $\cN=8$ supergravity U-duality $E_{7(7)}$  symmetry acts on  scalars and on  vectors of the theory.\footnote{This U-duality symmetry together with maximal supersymmetry is believed to be the reason why in perturbative $\cN=8$ supergravity the UV properties are better than expected naively,  and $\cN > 4$ supergravities may be even UV finite, if U-duality symmetry has no anomalies, see for example \cite{Beisert:2010jx,Kallosh:2018wzz,Gunaydin:2018kdz}.}
It was observed in  \cite{Ferrara:2016fwe} that one can start with M-theory in d=11 with its maximal supersymmetry, $\cN=1$, and compactify this theory on a specific $G_2$ manifold. Alternatively, one can start with type IIB  string theory in d=10 with its maximal supersymmetry, $\cN=2$, and compactify it on a $T^2\times T^2\times T^2$ manifold, or one can start directly in d=4 with its  maximal supersymmetry,  $\cN=8$.
In all these cases one ends up with a theory which depends on 7 complex scalar fields, each being a coordinate of the hyperbolic disk, 
\be
{\cal L}_{\rm kin} =- \sum_{i =1}^7 {\partial Z_i \partial \bar Z_i\over (1-Z_i\bar Z_i)^2} \ .
\label{Lkin} \ee
For each disk we have $3\alpha_i=1$. As long as all maximal supersymmetry is preserved, there is no potential in ungauged supergravity. 

The origin of the 7 hyperbolic disks is easiest to explain in the case of $\cN=8$ supergravity in d=4, which has a duality symmetry $E_{7(7)}$. For M-theory and string theory the explanation is available in  \cite{Ferrara:2016fwe}.
$\cN=8$ supergravity in d=4,  has  duality symmetry  $E_{7(7)}$. When the maximal $\cN=8$ supersymmetry is broken to the minimal $\cN=1$ supersymmetry, one finds a decomposition into 7 hyperbolic disks,
\be
E_{7(7)} \mathbb{R}\supset [SL(2, \mathbb{R})]^7 \ .
\ee 
since $[SL(2, \mathbb{R})]^7 $ is a subgroup of $E_{7(7)}$.
The corresponding kinetic terms are shown in eq. \rf{Lkin}.
To view this set of kinetic terms as a viable model of a single-field inflation, one can proceed by  cutting/identifying some of the moduli, so that the resulting kinetic term of a single disk becomes 
\be
{\cal L}_{\rm kin} = -3\alpha {\partial Z \partial \bar Z  \over (1-Z\bar Z)^2} \ , \qquad  3\alpha=7,6,5,4,3,2,1 \ ,
\label{7disk} \ee
replacing 7 units size disks,  each with $3\alpha_i=1$, as proposed in  \cite{Ferrara:2016fwe}.

Later, in \cite{Kallosh:2017ced} a dynamical mechanism, which we called `disk merger',  was proposed,  replacing the   cutting/identifying  moduli procedure suggested in   \cite{Ferrara:2016fwe}.   The choice of the potential (\K\, and superpotential) depending on 7 complex scalars was found with unbroken minimal $\cN=1$ supersymmetry, which dynamically either removes some of the disks, or identifies some of them with each other. All possibilities were listed, and the result  confirmed earlier kinematic choices made in \cite{Ferrara:2016fwe} and shown in Fig.~\ref{7disk2}.

All models with the potentials for the inflaton field preserved the kinetic terms originating from the 7 disks of M-theory/string theory/maximal supergravity. The result is a single disk kinetic term, which can only   take the 7 values above as shown in   \rf{7disk}. Now we have to remember that all observables in $\alpha$-attractors depend only on the kinetic geometric terms for the scalars, not on a choice of the inflaton potential. Therefore, the predictions from maximal supersymmetry spontaneously broken to minimal supersymmetry depend on the choice of $N$ so that 
our 7 benchmarks are
\be
n_s= 1- {2\over N}\ , \quad r=   {28   \over N^2} \, ,  {24   \over N^2} \, , 
 {20   \over N^2} \, , 
 {16   \over N^2} \, , 
 {12   \over N^2} \, , 
 {8   \over N^2} \, , 
 {4   \over N^2}   \ .
\label{BM}\ee

When our models are derived from maximal supersymmetry, spontaneously broken to the minimal one, we do not have anything above $\alpha = 7/3$ and below $\alpha = 1/3$. The $\alpha$-attractor realizations of the potential of the Starobinsky model, and of the Higgs inflation potential, is an intermediate one, $\alpha = 3/3=1$.

All 7 cases, which we show in  Fig.~\ref{7disk2}  are testable and falsifiable if B-modes are not detected at $r\gtrsim 10^{-3}$ since for $\alpha=1/3$ and the largest value of $N=60$ we find
\be
r=    {4   \over N^2} :  \qquad r=    {4   \over  60 ^2}\approx 10^{-3} \ .
\ee
The earliest U-duality benchmark starts with the 7 disk merger, with $3\alpha=7$ at smallest value of $N = 50$ 
\be
r=    {28  \over N^2} :  \qquad r=    {28   \over  50^2}\approx 10^{-2} \ .
\ee
Therefore these models compatible with the data are also compatible with $\cN=8$ supergravity (and M-theory and string theory) have the following property: $U$-duality and maximal  supersymmetry lead to
\be 
10^{-3} \lesssim r \lesssim 10^{-2} \ .
\ee

In the range $ 10^{-3} \lesssim r \lesssim 10^{-2}$ 
there are U-duality benchmarks for the B-mode detection, based on the simplest T-model which is shown by two yellow straight lines on $n_s-r$ plot in the Fig.~\ref{7disk2}. Note  that all $\alpha$-attractor models in the  small $\alpha$ limit asymptotically  give the same predictions as the simplest T-model, as we show in Fig.~\ref{alpha1}. The region $ 10^{-3} \lesssim r \lesssim 10^{-2}$  is still  pre-asymptotic. The spread of lines  shown in Fig.~\ref{alpha1} in the left panel is small in this region  for more general T-models. However, in the right panel in Fig.~\ref{alpha1}, one can see that the general class of E-models has a certain shift  towards   higher values of $n_s$, as compared to the T-models.

Therefore a possible detection of the B-modes in the range $ 10^{-3} \lesssim r \lesssim 10^{-2}$  to the right of the two yellow lines in Fig.~\ref{7disk2} could be associated with the predictions of the E-models in this region, as we show in Fig.~\ref{alpha1}. For these E-models, U-duality origin of the kinetic term is intact. The slight shift of the benchmarks to the right originates from the slight dependence of the predictions of the theory on the choice of the potential, since we are in the region where E-models have not reached the attractor point yet.

\subsection{Special benchmarks}\label{sb}
All $3\alpha=7,6,5,4,3,2,1$ are on equal footing with regard to the origin of their kinetic term from theories with maximal supersymmetry. However, few of these have an additional meaning.
\begin{itemize}
  \item the predictions of $3\alpha=6$, $\alpha =2$ theory are known to be the same as in the string theory model of fibre inflation \cite{Cicoli:2008gp,Kallosh:2017wku}. This is the second from the top benchmark in Fig.~\ref{7disk2}.
  \item the predictions of $3\alpha=3$, $\alpha=1$  theory are known to be the same as in Starobinsky model ~\cite{Starobinsky:1980te} and in Higgs inflation model~\cite{Salopek:1988qh,Bezrukov:2007ep} and in conformal inflation model~\cite{Kallosh:2013hoa}. This is the third  from the bottom benchmark in Fig.~\ref{7disk2}.
  \item The case $3\alpha=2$, $\alpha= {2\over 3} $ has the interesting property that the moduli space curvature $|\mathcal{R}|= {2\over 3\alpha}=1$. It is also one of the  possible candidates for the characteristic  scale of the potential \cite{Linde:2016hbb}. This is the second  from the bottom benchmark in Fig.~\ref{7disk2}.
    
    \item The case $3\alpha=1$, $\alpha={1\over 3}$  is special. First, it is the last one in the 7-disk story  \cite{Ferrara:2016fwe,Kallosh:2017ced} which follows from U-duality and maximal supersymmetry:  M-theory in d=11, string theory in d=10, and maximal supergravity in d=4. Moreover,  kinetic term with $\alpha={1\over 3}$, a unit disk geometry,  can also be derived from the maximal superconformal theory in d=4, as we explained in \cite{Kallosh:2015zsa} and  in Appendix C of this paper. This is the last  from the bottom benchmark in Fig.~\ref{7disk2}.

  \end{itemize}

\section{D-brane inflation models}\label{dbranes} 

The string theory origin of D-brane inflation model is associated with the  KKLMMT model~\cite{Kachru:2003sx}, where D3-brane-$\overline {\rm D3}$-brane interaction was studied in the context of the volume modulus stabilization. Earlier proposals for D-brane inflation relevant to our current discussion  were made in~\cite{Dvali:2001fw,Burgess:2001fx,GarciaBellido:2001ky}. 
 
D-brane  inflation models have a potential  \cite{Planck:2018jri} 
\be
V_{\rm D-brane}\sim 1-  {m^{k}\over \phi^{k} } +\cdots
\label{htdb}\ee
Here the ellipsis stay for higher order terms. 
These models were studied in detail in  \cite{Lorenz:2007ze,Martin:2013tda} and in Planck 2013 \cite{Planck:2013jfk}. In the small $m$ limit,  predictions of these models for $n_{s}$ do not depend on $m$ and on the omitted higher order terms, i.e. they exhibit an attractor behavior:
 \be
n_{s}= 1- {2 \over   N} {k+1\over k+2} \ .
\label{htdbns}\ee
D-brane inflation models with the potential  ignoring higher order terms  are called BI models (from brane inflation)  \cite {Martin:2013tda}. 
\be
V_{BI} \sim 1- {m^{k}\over \phi^{k} } \ .
\label{BI} \ee
  This potential is unbounded from below, so it does not describe a consistent cosmological evolution.

Consistent generalizations of these models were  proposed in \cite{Kachru:2003sx} in the context of D3 brane inflation. These models  were generalized and   called KKLTI  (KKLT inflation)  in  \cite{Martin:2013tda} and further developed in  \cite{Kallosh:2018zsi}. They have potentials
\be
V_{KKLTI} \sim   {\varphi^k\over m^k + \varphi^k}= \Big( 1+  {m^k\over \varphi^k }\Big )^{-1}\, , \quad k=7- p \ .
\label{KKLTI}\ee
The derivation of these potentials following \cite{Kachru:2003sx,Martin:2013tda} involves the inverse harmonic function for  Dp-brane potentials in Euclidean $9- p$ dimensions.   Note that the Dp-brane potentials of this type can contribute to the 3d vacuum-like potential energy density and lead to   inflation only for $p \geq 3$, i.e. for $k \leq 4$. 

The cosmological evolution of these models   was described in detail in  \cite {Martin:2013tda,Lorenz:2007ze}, and studied more recently in \cite{Kallosh:2018zsi,Kallosh:2019jnl}. At very large   $m$ they have the same predictions as the  models with $V\sim \varphi^k$, but at $m \ll 1$ and  and $r \lesssim 10^{-3} $ they have the same predictions \rf{htdbns} as the BI models. 
For example,
for a quartic KKLTI model, $k=4$, $\beta = {5\over 3}$ for D3-branes with $ p=3$,  we find for small $m$ that
\be
 n_s \approx  1- {5 \over  3 N}   \, , \qquad  r\approx {4 m^{{4\over 3} }\over (3N)^{5\over 3}} \, .
 \label{quartic}  \ee
For the quadratic KKLTI model, $k=2$, $\beta= {3\over 2}$ for D5-branes with $ p=5$,  we find for small $m$ that
\be
 n_s \approx  1- {3 \over  2 N}\ ,  \qquad  r\approx   {\sqrt 2\, m \over N^{3\over 2}}   \ .
 \label{quadratic}\ee
 \begin{figure}[!h]
\begin{center}
\includegraphics[width=6cm]{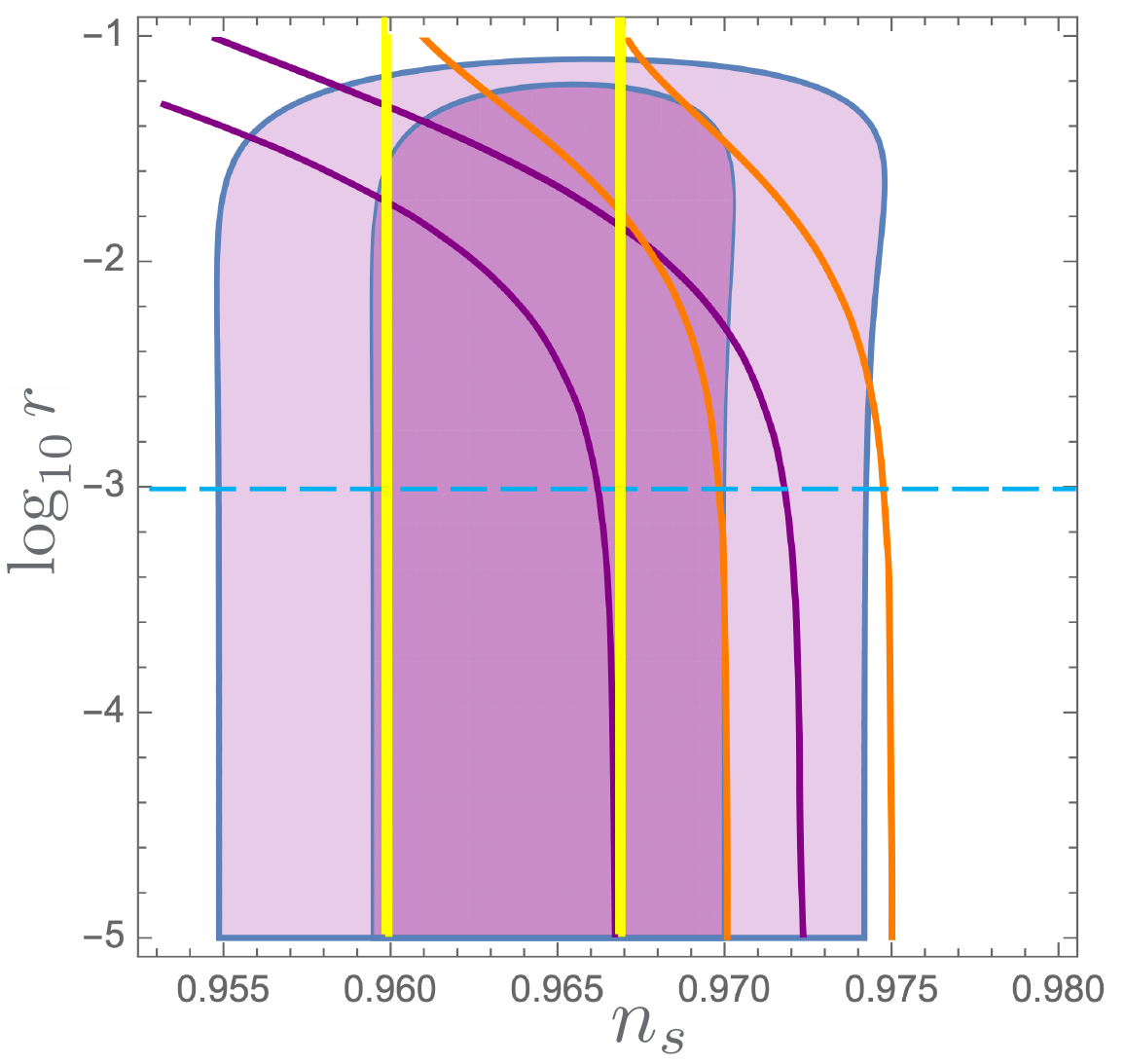}
\caption{\footnotesize Predictions of $\alpha$-attractors  and KKLTI models.   Two yellow lines  are for the quadratic T-model of $\alpha$-attractors at $N = 50$ and $N = 60$.  Two purple lines are for the quartic KKLTI model, two orange lines  show the predictions of the quadratic KKLTI model. Predictions of all of these models converge to their asymptotic values for $r \lesssim 10^{{-3}}$ indicated by the blue dashed line.
 } 
\label{QQ}
\end{center}
\vspace{0cm}
\end{figure}

As one can see from Fig.~\ref{QQ},   quartic and quadratic versions of D-brane inflation in Fig.~\ref{QQ} with a non-singular potential bounded from below are among the simplest string theory motivated models to be tested by the  B-mode searches. Their predictions for $n_{s} $ in the small $m$ limit are positioned  to the right  of the $\alpha$-attractors in the $n_s-r$ plane. At small $r$, the  combination of these models describes $\beta$ stripes with $\beta = 2, \, {5\over3}$, and ${3\over 2}$. A combination of these three classes of models almost completely covers the area in the ($n_{s}$, $r$) space favored by Planck 2018 at the $1\sigma$ level.

  There are no specific predictions for the value of the parameter $m$ in these models. If anything, presently available string theory examples of this kind have $m \ll 1$ and 
$r\approx 10^{-6}-10^{-10}$, as suggested in the discussion of the KKLMMT model~\cite{Kachru:2003sx} in Appendix C, and in examples in \cite {Martin:2013tda}.  However, we are at  early stages of development of such models, so it might be possible to have greater values of $r$ in such models. There are no obvious constraints on $m$ and $r$ in the string theory motivated supergravity versions of these models \cite{Kallosh:2018zsi}.     Moreover, the potentials described in this section can be obtained in other way. For example, a quadratic model $V  \sim   {\varphi^2\over m^2 + \varphi^2}$ was proposed in \cite{Dong:2010in} as an example of a flattening mechanism due to the inflaton interactions with heavy scalar fields. In the next section we will show that similar potentials can be also obtained in the context of pole inflation;  these models do not impose any constraints on the values of $m$ and $r$.

 \section{General Dp-brane and pole inflation models with \boldmath{$q \not =2$}}\label{general}
 
\subsection{Pole inflation models with \boldmath{$q \not =2$}}

 The hyperbolic two-dimensional geometry naturally leads to a pole inflation with a pole of order $q=2$ and the residue of the pole defined by the {\it \K\, curvature of the complex  manifold}. Meanwhile, the one dimensional slice of the geometric scalar manifold, where the sinflaton partner of the inflaton is stabilized, corresponds to a one dimensional Riemannian manifold of the form
 \be
 ds^2= g_{\rho\rho}(\rho) d \rho^2 = d\varphi^2 \ , \qquad g_{\rho\rho}(\rho) =  {1\over 2} {a_q\over \rho^q} \ ,
\label{1d} \ee
which we encounter in the pole inflation models of Sec. \ref{alpha}.
 In the case that $q=2$ the \K\, manifold metric of the Poincar\'e disk
\be
g_{Z\bar Z}=  {3\alpha \over (1-Z\bar Z)^2}
\ee
 is given in eq. \rf{kinZ} and therefore $a_{2} \equiv {2\over 3\alpha}$ is associated with the \K\, manifold curvature $a_{2} = - \mathcal{R}_K$. The same curvature is obtained using the half-plane coordinates with the metric 
\be
g_{T\bar T} = { 3\alpha \over (T+\bar T)^2} \ ,
\ee
as given in eq. \rf{half}.

Note that a one-dimensional Riemannian manifold of the kind shown in eq. \rf{1d} does not have an intrinsic curvature, it is always locally isometric to a straight   line. Indeed, a local change of variables from $\rho$ to $\varphi$ leads to a trivial metric $g_{\varphi\varphi}=1$.\footnote{Note that globally such a change of variables may not be well defined, which may lead to confusing situations, some of which will be discussed below.} Formally, the Riemann curvature tensor has  a single component $R_{1111}$, but this element is required to be equal to $0$  due to  the anti-symmetric property of the Riemann curvature tensor $R_{ijkl}$ under the interchange of the indices.

Therefore interpretation of the single field pole inflation with $q\neq 2$ may rise some questions because there is no geometric structure associated with the one-dimensional pole-type metric. Nevertheless this framework is quite  useful since it provides a simple unified interpretation of a rather large class of cosmological attractors, including hilltop models and D-brane inflation.

Following our discussion in section \ref{alpha}, we start with the Lagrangian \rf{action},
 \be
{\cal L} = {\cal L}_{\rm kin} - V  = - {1\over 2} {a_q\over \rho^q} (\partial \rho)^2- V(\rho) \ .
\label{pole3}\ee
For $q=2$ this theory describes $\alpha$-attractors. Consider now the case  $q \not =2$. Then the canonical variable $\vp$ is related to $\rho \geq 0$ as follows:
\be
\rho =\Bigl |{(2-q)\vp\over 2 \sqrt {a_q}}\Bigr |^{2\over 2-q}  \ .
\ee
The absolute value appears because the equation to be solved is $\pm {\partial \rho\over \rho^{q/2} }= \phi$, with one of the signs, and the solution for $\rho$  must be positive. However, already at that stage the situation becomes somewhat delicate.

For $q = 2$, the solution of the corresponding equation is $\rho = e^{-\sqrt {a_{2}}\varphi} = e^{-\sqrt {2\over 3\alpha}\varphi}$. This transforms the field $\rho$ in the original range of values $0< \rho < +\infty$ into a canonical field $\vp$ defined in the full unconstrained range $-\infty < \vp < +\infty$. By solving equations of motion using $\varphi$ variables, one can never reach infinitely large values of $\varphi$ within finite time. Therefore the same should be true for the field $\rho$: neither the values of $\rho = +\infty$ nor the point $\rho = 0$ can be reached within finite time.

Meanwhile for $q < 2$, the canonical field $\vp$ can reach the point $\vp = 0$  within finite time, and therefore the point  $\rho = 0$ is also accessible. This means that the coordinate system covering $0< \rho < \infty$ is incomplete; see  \cite{Karamitsos:2019vor} for a related discussion.

This is not necessarily a problem. For example, for   $q = 3/2$ one has $\rho = {\vp^{4}\over 16 a_q^2}$. Therefore the potential $V(\rho)$ becomes a certain function of  or $\vp^{4}$, which can be easily continued to $\vp < 0$, thus making it possible to consider the full range of values $-\infty <\vp < +\infty$. This simple procedure works for $q= 2\mp1/n$, resulting in $\rho \sim \vp^{\pm 2n}$. One can use a similar procedure for general values of $q < 2$, as long as it does not lead to anomalous behavior of $V(\vp)$ at $\vp = 0$.  

For $q > 2$, the transformation to canonical variables makes the change $\rho = 0 \to \vp = +\infty$ and  $\rho = +\infty \to \vp = 0$. Thus the field cannot reach the singularity at $\rho = 0$, which would require an infinitely long journey for the canonical field $\vp$, but it can reach $\rho = +\infty$, which is, therefore, not a true physical infinity. A potential  $V(\rho)$ slowly growing when $\rho$ approaches $0$ looks like an infinite  plateau at large $\vp$, but one should check whether the potential has an acceptable behavior (e.g. whether it is differentiable) at $\vp = 0$, prior to performing its continuation to $\vp < 0$.

Let us consider  some simple potentials, and check how they look in canonical variables.  
\be\label{ch1}
V(\rho)  = V_{0}(1-c\, \rho) \Longrightarrow   V_{0}\left(1-  \Bigl( {\vp\over m}\Bigr)^{{2\over  2-q}}\right) \ , 
\ee
where 
\be
m = {2\sqrt {a_q} \over |2-q| \, c^{2-q\over 2}} \ .
\ee
In particular, for $q= 1$ one has a quadratic hilltop potential $V = V_{0}(1-   {\vp^{2}/m^{2}})$.
For $q= 3/2$ one has a quartic hilltop potential $V = V_{0}(1-   {\vp^{4}/m^{4}})$.
For $q= 5/2$ one has a quartic BI D-brane potential  $V = V_{0}(1-   {m^{4}/\phi^{4}})$.
For $q= 3$ one had a quadratic BI D-brane potential 
$V = V_{0}(1-   {m^{2}/\phi^{2}})$.
As we already discussed, all of these potentials are unbounded from below, and therefore they should be discarded. Nevertheless, it is remarkable that all of them can be  easily obtained from the simple linear potential $V_{0}(1-c\, \rho)$ in the context of pole inflation.
 
One can improve these potentials and make them positively definite by doing the same as we did in the derivation of the Starobinsky model, or the E-models in the context of $\alpha$-attractors. One can introduce the positively defined potential
 \be\label{genE}
V(\rho)  = V_{0}(1-c\, \rho)^{2} \Longrightarrow  V_{0}\left(1-  \Bigl( {\vp\over m}\Bigr)^{{2\over  2-q}}\right)^{2} \ . 
\ee
For $q = 1$ one has the usual Higgs-type potential, 
  \be
V = V_{0}\Bigl(1-   {\vp^{2}\over m^{2}}\Bigr)^{2} \ . 
\ee
which leads to inflation for $m \gg 1$  \cite{Linde:1994hy,Linde:2007fr,Kallosh:2007wm}, but its predictions do not provide a particularly good fit to the Planck 2018 data. For $q =3/2$ one has the squared hilltop potential introduced in our previous paper \cite{Kallosh:2019jnl},
  \be\label{44}
V = V_{0}\Bigl(1-   {\vp^{4}\over m^{4}} \Bigr)^{2} \ . 
\ee
This theory   provides a good fit to the Planck 2018 data \cite{Kallosh:2019jnl}, but previously it did not have any physical or mathematical motivation.  Now we see that this model is a generalization of the  E-models with $q = 2$  for pole inflation with $q =3/2$. However, the attractor nature of this model is not helpful here since this hilltop model in the attractor regime with $m < 1$ predicts too small values of $n_{s}$.  This model is compatible with the Planck 2018 data only for $m \gg 1$, where its predictions are very different from the predictions of the simple hilltop inflation models $V = V_{0}(1-   {\vp^{4}/m^{4}})$  \cite{Kallosh:2019jnl}.

For $q =7/4$ one has the squared hilltop potential introduced in our previous paper \cite{Kallosh:2019jnl}.
  \be \label{88}
V = V_{0}\Bigl(1-   {\vp^{8}\over m^{8}} \Bigr)^{2} \ . 
\ee
This theory provides only a marginal fit to the Planck 2018 data  in the attractor regime with $m < 1$, where it  predicts  $n_{s}= 1-{7\over 3N}$. This value  is  smaller than the $\alpha$-attractor prediction  $n_{s}= 1-{2\over N}$ by about $0.006$.

One may study a more complicated possibility:
\be
V(\rho)  =  V_{0}\, e^{-c\, \rho} \Longrightarrow V_{0}\, e^{- \bigl( {\vp\over m}\bigr)^{{2\over  2-q}}} \ . 
\ee
These potentials, in different notation, were used in \cite{Mukhanov:2013tua,Abazajian:2016yjj}. Note that these functions for $q< 2$ look like the hilltop models with potentials that do not have a minimum, but instead become super-exponentially small at large $\phi$. They might be useful for a description of quintessential inflation. For $q > 2$, these functions describe plateau potentials with an extremely flat minimum, such that the mass of the field vanishes at $\vp = 0$.

Until now,  we assumed that when the field approaches $\rho = 0$, the potential grows linearly.  However, one may also consider the models where the potential $V(\rho)$ is symmetric with respect to the change $\rho \to -\rho$. This would mean that the simplest potential $V(\rho)$ at small $\rho$ is 
\be
V(\rho)  = V_{0}(1-c^{2}\, \rho^{2})+... \Longrightarrow   V_{0}\left(1-  \Bigl( {\vp^{2}\over m^{2}}\Bigr)^{{2\over  2-q}}\right)+... \ . 
\ee
For example, for $q= 1$ one has a quartic hilltop potential $V = V_{0} (1-   {\vp^{4}/m^{4}})$. Meanwhile in the theory \rf{ch1} with a linear term in the potential $V(\rho)$ the resulting potential for $q = 1$ was $V = V_{0} (1-   {\vp^{2}/m^{2}})$.
This illustrates a distinctive feature of the attractors with $q \not = 2$. For $\alpha$-attractors ($q = 2$), elimination of the linear term in the potential $V(\rho)$ changes $\alpha$, and therefore $r$, but it does not affect $n_{s}$. Meanwhile for $q \not = 2$ elimination of the linear term in the potential $V(\rho)$ changes $n_{s}$, moving the predictions from one $\beta$ stripe to another.

\subsection{D-brane inflation and pole inflation   with \boldmath{$q >2$}}

By construction, all pole inflation models are attractors. Here we will consider a subclass of these models closely related to D-brane inflation.
As we already mentioned, a simplest representative of this class is the quartic BI inflation potential   $V = V_{0}(1-   {m^{4}/\phi^{4}})$, which correspond to the theory with $q= 5/2$. This potential is unbounded from below, but one can consider some of its consistent generalizations,
such as  the potential
\be
V(\rho)  = {V_{0}\over  1+c\, \rho} \Longrightarrow  V_{0}\left(1+  \Bigl( {\vp\over m}\Bigr)^{{2\over  2-q}}\right)^{-1} \ . 
\ee
For $q = 5/2$ one finds the  KKLTI $D_{3}$-brane inflation potential
   \be\label{quart}
V = V_{0}\Bigl(1+   {m^{4}\over \vp^{4}} \Bigr)^{-1} = V_{0} \,     {\vp^{4}\over \vp^{4}+m^{4}} \ . 
\ee
For $q = 8/3$ one finds the KKLTI $D_{4}$-brane inflation potential
\be\label{cub}
V = V_{0}\Bigl(1+   { m^{3}\over \vp^{3}} \Bigr)^{-1}  = V_{0} \,     {\vp^{3}\over \vp^{3}+m^{3}} \ . 
\ee
For $q = 3$ one finds the KKLTI $D_{5}$-brane inflation potential
   \be\label{quadr}
V = V_{0}\Bigl(1+   { m^{2}\over \vp^{2}} \Bigr)^{-1}  = V_{0} \,     {\vp^{2}\over \vp^{2}+m^{2}} \ . 
\ee
Finally, for $q = 4$ we have the KKLTI $D_{6}$-brane inflation potential
   \be\label{lin}
V = V_{0}\Bigl(1+   { m\over \vp} \Bigr)^{-1}  = V_{0} \,     {\vp\over \vp+m} \ . 
\ee
The potentials \rf{cub} and \rf{lin} could seem unphysical because they become negative and are unbounded from below for $\vp < 0$. However, this is not a real problem. In the D-brane context, $\vp$ is a measure of the distance between the branes, which is positive. One can describe the theory by a potential symmetric with respect to the change $\vp \to -\vp$, using the procedure discussed in the previous subsection, i.e. effectively replacing $\phi$ by the positive distance $\sqrt{\phi^{2}}$.

Now we are using several different parameters, closely related to each other: $p$, $k$, $\beta$ and $q$. Relations between these parameters for the 4 different Dp-brane models discussed above is explained in the Table 1:

{\footnotesize
\begin{table}[h]
\begin{center}
\label{topotable}
\begin{tabular}{|c||c|c|c| }
\hline ~p~ &  ~k~& ~$\beta$~  & ~q~  \\
\hline \hline
\hline $3$  &   4 & 5/3 & 5/2
 \\
\hline $4$  &  3 &8/5  &8/3  
\\
\hline $5$  &  2 &3/2 & 3
 \\
\hline $6$  &  1 &4/3 & 4
 \\
\hline
\end{tabular}
\caption{\footnotesize Relation between the parameters $p$, $k$, $\beta$ and $q$ for the Dp-brane inflation models.}
\end{center}
\end{table}
}

Thus the potentials mentioned above have two independent interpretations, as the D-brane inflation potentials, and as potentials of the cosmological attractors in the context of the pole inflation.
This can be very useful for interpretation of the models with $m > 1$. Indeed,  as we already mentioned, the  parameter $m$ in D-brane inflation typically is very small,  $m\ll 1$, see~\cite{Kachru:2003sx,Martin:2013tda,Lorenz:2007ze}. Meanwhile there is no such constraint for general pole inflation models. In other words, the models \rf{quart}-\rf{lin} represent  consistent pole inflation attractors for any $m$, independently of their string theory interpretation. Note that these potentials are symmetric with respect to the change $\phi \to -\phi$, and in this sense they are similar to the T-models of $\alpha$-attractors.
The predictions of this set of models, in combination with the simplest T-models and E-models of $\alpha$-attractors, are shown in Fig. \ref{6pots}   in the Introduction. 

\begin{figure}[!t]
\begin{center}
\hskip 0.35cm\includegraphics[width=6cm]{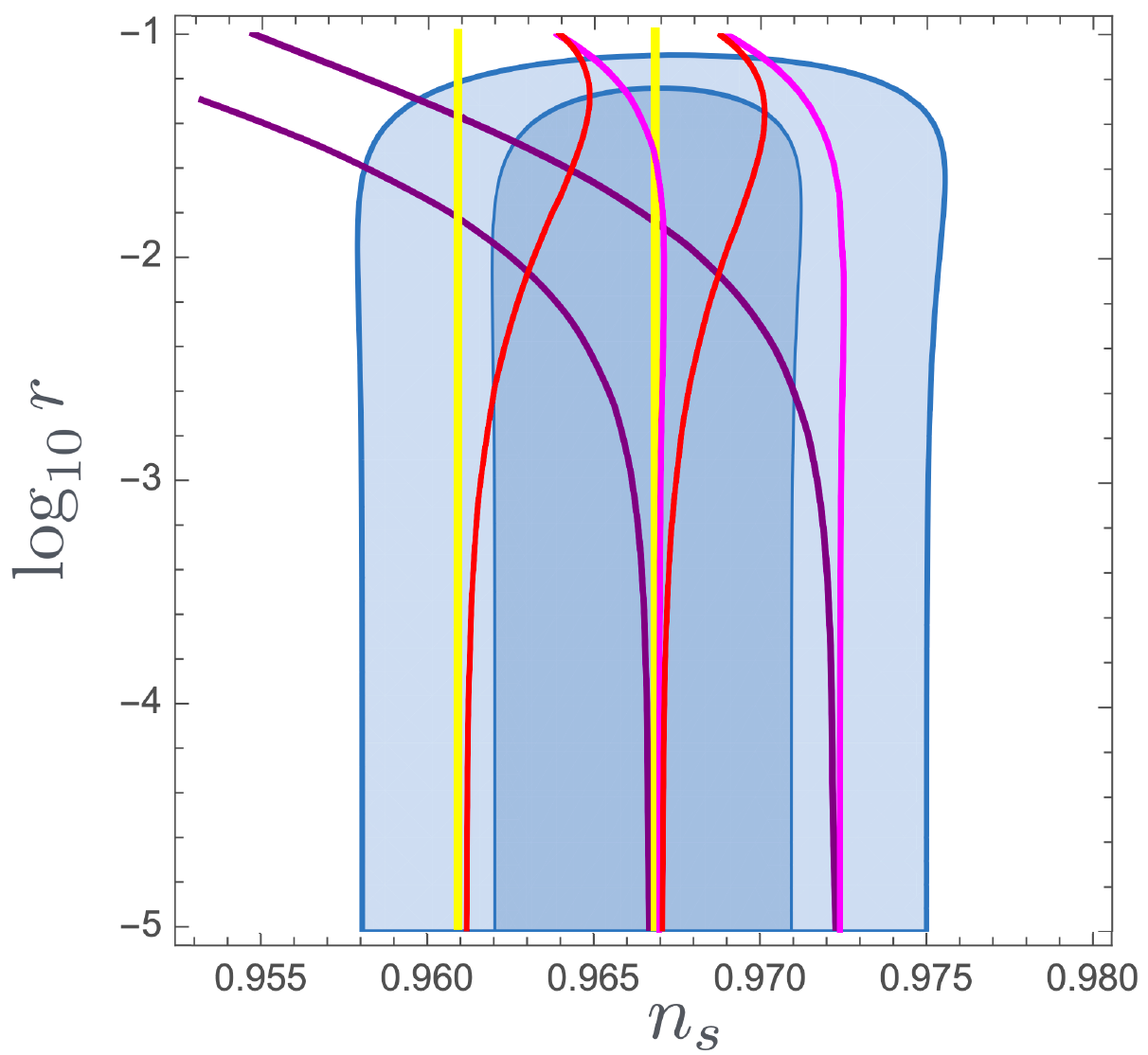}
\caption{\footnotesize Predictions of pole inflation models with $q = 2$ ($\alpha$-attractors) and $q = 5/2$. Two yellow lines  shows predictions of the simplest T-models for N = 50, 60. Two  red lines show predictions of the simplest E-models.   Two purple lines are for the quartic KKLTI model \rf{quart}, two magenta lines  show the predictions of the model \rf{quart2}. The set of these simple $q = 2$ and $q = 5/2$ attractors completely covers the sweet spot of the Planck 2018 data. 
 } 
\label{QQ2}
\end{center}
\vspace{0cm}
\end{figure}

Another set of consistent pole inflation models   is described by the potential \rf{genE} with $q > 2$, generalizing $E$-models. In particular, 
for $q = 5/2$ one has a potential 
  \be\label{quart2}
V = V_{0}\Bigl(1-   {m^{4}\over \vp^{4}} \Bigr)^{2}   \ . 
\ee
It is singular at $\vp = 0$, and it has a minimum at $\vp = m$. It is instructive to show the predictions of the simplest   $\alpha$-attractors ($q = 2$) simultaneously with the predictions of the $q = 5/2$ attractors which we just discussed, see Fig. \ref{QQ2}. As one can see from this figure,  the set of the  simplest $q = 2$ and $q = 5/2$ attractors completely covers the  dark blue $1\sigma$ area of the Planck 2018 data for $n_{s}$ and $r$. 

Moreover, a combination of the T-models  (yellow lines) and the model \rf{quart2} (magenta lines) is already sufficient to cover the sweet spot of the Planck data. In the small $r$ limit, any single choice of the family of $\alpha$-attractors  (T or E-models) in combination with any family of attractors with $q = 5/2$ form two stripes which are  sufficient to describe the presently available data.


\section{On phenomenological parametrizations of the CMB data}\label{Char}
 
In our investigation we were trying to identify the simplest inflationary models  motivated by fundamental physics, which would provide a good match to the Planck data. All models that we have  analyzed in this paper have an unusual property: in the small $r$ limit, their predictions form vertical $\beta$ stripes in the ($n_{s}$, $r$) space,
\be\label{ourbeta}
 1-n_s =  {\beta\over N} \ .
 \ee
Few years ago, this property could seem very unusual indeed. Among dozens of  models analyzed in {\it   Encyclop\ae dia Inflationaris} \cite{Martin:2013tda}, which was written prior to the invention of $\alpha$-attractors, only two models possess this property: hilltop inflation and D-brane inflation, in the limit $m \ll 1$.

An alternative approach developed in   \cite{Mukhanov:2013tua,Creminelli:2014nqa,Abazajian:2016yjj} is to be agnostic with respect to fundamental physics, postulate the equation \rf{ourbeta},
\be
1-n_s(N) =  {\beta\over N} \ ,
\label{tilt}\ee
up to sub-leading corrections in an expansion in  $1/N$,  and then study consequences of this hypothesis.  (We  replaced  $p+1$ used in this equation in 
\cite{Abazajian:2016yjj} by $\beta$ to avoid confusion with the Dp-brane notation.)

The proposal made in  \cite{Creminelli:2014nqa,Abazajian:2016yjj} is to solve eq. \rf{tilt}. The solution of this equation for the slow-roll parameter $\epsilon$ is 
\be
\epsilon (N) = {\beta-1\over 2N} {1\over 1\pm (N/N_{\rm eq} ) ^{\beta-1} } \ .
\label{ep}\ee
Here $N_{\rm eq}$ is an integration constant. 
If $N_{\rm eq} \ll N$, which is expected at sufficiently small $r$, this can be brought to the form
\be
\epsilon (N) =   {(\beta-1)  N_{\rm eq}^{\beta -1} \over 2 N^{\beta} }\, , \label{ep2a}\ee
which yields 
\be
r (N, \beta, N_{\rm eq}) = 16 \epsilon (N) = { 8 (\beta-1)  \, N_{\rm eq}^{\beta-1} \over  N^{\beta} } \ .   \label{ep2}\ee
Since the  models with $n_s(N) =1 - {\beta\over N}$ studied in   \cite{Mukhanov:2013tua,Creminelli:2014nqa,Abazajian:2016yjj} can be also obtained in the context of pole inflation,   we can compare the expression for $r$ obtained   in eq. \rf{pole}, 
\be
 r= 8 \,   a_q^{\beta-1} \Big ({{\beta-1}\over {N} }\Big )^{\beta} \ ,
\ee
with the expression for $ r (N, \beta, N_{\rm eq})$  in  \rf{ep2}. This gives 
\be\label{neqq}
N_{\rm eq}= (\beta-1) a_q  = {a_q\over q-1}\ .
\ee
 Here $\beta={q\over q-1}$, $q$ is the order of the pole in eq. \rf{action}, and $a_q$ is the residue, redefined to absorb the arbitrary constant $c$ in the potential, as explained in section 2.
 
Thus, the theory of pole inflation provides a simple interpretation of the   important but somewhat obscure parameter $N_{\rm eq}$ in terms of   the residue at the pole.

The next step in  \cite{Abazajian:2016yjj} is the introduction of the characteristic scale of inflationary potential $M$. The definition the characteristic scale implicitly used in \cite{Abazajian:2016yjj} is\footnote{R. Flauger, private communication} 
\be
 M^2 = {4\over \beta-1}N_{\rm eq} \ .
\label{Neq}\ee
However, Eq. \rf{Neq} is  a definition of a new concept, the characteristic scale $M$, which was not explained and explicitly presented in \cite{Abazajian:2016yjj}. 
It is interesting therefore that Eq. \rf{Neq}, up to the coefficient $4$,  has a simple interpretation in terms of pole inflation. Indeed, Eq. \rf{Neq}, in combination with \rf{neqq}, yields
\be
 M^2 = 4\, a_q \ .
\label{aq}\ee

Note that  the choice of the coefficient $4$ in \ref{Neq}) and in front of $a_{q}$ in  equation \rf{aq} is just a matter of preference or convenience.
Since all equations in this section apply to models with arbitrary values of $q$ and $\beta$, it may be better to use a definition of the characteristic scale motivated by the  theory of $\alpha$-attractors, where $\beta = q = 2$ and  $a_q =  {3\alpha\over2}$.
If instead of  $M^2 = 4\, a_q$ we identify the square of the characteristic  scale $M^{2}$ with the pole residue $a_q$, 
\be\label{M1}
M^2 =  a_q \ ,
\ee
we will have 
\be
M = \sqrt{3\alpha\over2} \ ,
\ee
and the asymptotic behavior of the  inflaton potential for all $\alpha$-attractors will be described by a simple intuitively appealing equation 
\be
V = V_{0}\Bigl(1- e^{-\vp/M}+O\bigl(e^{-2\vp/M}\bigr)\Bigr) \ ,
\ee
Another advantage of this definition, used in \cite{Linde:2016hbb,Kallosh:2019eeu}, is that $M$ would have a simple interpretation in terms of the  negative curvature of the hyperbolic moduli space  \cite{Ferrara:2013rsa,Kallosh:2015zsa}
\be
 \mathcal{R}= - M^{-2} \ .
\ee
Now we have an additional argument in favor of this definition: $M^{2}$ would coincide with the residue at the pole $a_q$.

Yet another physically interesting definition would be 
\be\label{M3}
M^2 = 2 a_q \ ,
\ee
 which would imply 
\be\label{ME}
M = \sqrt{3\alpha} \ .
\ee
In that case, $M$ would be given by the radius of the Escher disc discussed in section  \ref{proposal}. At $N=60$ it would mean that for $\alpha$-attractors the Planckian value of this characteristic scale is $r= {4\over 60^2} \approx  10^{-3}$. This value of $r$ would also coincide with the lowest U-duality benchmark for $\alpha$-attractors, see the black line in Fig.~\ref{7disk2} and a discussion in Appendix \ref{conf}.

If we are planning to use the results of the B-mode search for finding the characteristic scale of inflation \cite{Abazajian:2016yjj}, it would be important to find  the best physically  motivated definition of this quantity. For example, the difference by the factor of 2 between the possible definitions of $M$  for $\alpha$-attractors given in \rf{aq} and \rf{M1} leads to the change by the factor of 4 between the values of $r$ for $M = 1$.  Once we make a well motivated choice of $M$ for $\alpha$-attractors ($\beta = 2$), one can multiply it by any function $F(\beta)$ such that $F(2) = 1$. This modification is not required, but it may be useful if one wants to associate the Planckian characteristic scale $M=1$ with the same value of $r \sim 10^{{-3}}$ for models with all relevant values of $\beta$, e.g. in the range $2 < \beta < 5/2$.

\section{Summary }

 This investigation, involving also a series of our recent papers \cite{Kallosh:2018zsi,Kallosh:2019jnl,Kallosh:2019eeu},  started soon after the Planck 2018 data release \cite{Planck:2018jri}. The main goal of this paper is to conclude this series of investigations by developing a unified description of the models favored by Planck 2018.  

As we mentioned in the Introduction, Ref.  \cite{Planck:2018jri} described three different classes of models of such type. The first class of models includes the Starobinsky model, the Higgs inflation model, the GL model, and the large class of $\alpha$-attractors with potentials $V \sim   1 - e^{-\sqrt {2\over 3\alpha}\varphi}  + \ldots $, which  embedded and generalized all of these models \cite{Kallosh:2013yoa,Carrasco:2015uma,Kallosh:2017wnt}.  The second class includes   the hilltop inflation models  with potentials $V\sim  1-{\varphi^{k}\over m^{k}} +... $  \cite{Linde:1981mu,Boubekeur:2005zm}.
The third class of models favored by Planck 2018 includes Dp-brane inflation models with  $V\sim  1-{m^{k}\over \varphi^{k}}   + \dots   $   \cite {Martin:2013tda,Planck:2013jfk,Kallosh:2018zsi}.

These three classes of models have some similarity: For $m\ll 1$ ($\alpha \lesssim 1$), they have an attractor regime: their predictions for $n_{s}$ do not depend on the higher order terms in the inflaton potential, on  $m$ and on $r$:  $n_{s}= 1-{\beta\over N}$.  In the $\alpha$-attractors models one has $\beta =2$, hilltop models have $\beta = 2\, {k-1\over k-2}> 2$, and D-brane models have $\beta = 2\, {k+1\over k+2} < 2$.    

Despite these similarities, theoretical motivation and observational status of these models is very different. The hilltop inflation in the attractor regime is strongly disfavored by Planck 2018 data, unless one considers models with $k \gtrsim 7$. Outside the attractor regime, for $m \gtrsim 1$, one cannot neglect the higher order terms, which are necessary to avoid global collapse of the universe in such models.   We are unaware of any natural version of hilltop inflation that would make  predictions reproducing the green area in the Planck, CMB-S4 and PICO Figs. \ref{CMBS42019}, \ref{F00}; see a detailed discussion of this issue in \cite{Kallosh:2019jnl}.

As for $\alpha$-attractors, they have a compelling theoretical motivation in the context of supergravity, and their predictions easily cover the left hand side of the  $1\sigma$ area in the ($n_{s}$,~$r$) space favored by Planck 2018.  Meanwhile we found that the right-hand side of the $2\sigma$ area  is completely covered by predictions of models with phenomenological potentials   \rf{quart}-\rf{lin}, which can be associated with D-brane inflation. These predictions, shown in Fig. \ref{6pots} in combination with  the predictions of $\alpha$-attractors, form a series of stripes with $\beta= 2, \  {5\over 3}, \   {8\over 5}, \  {4\over 3}, \  {3\over 2}$. The last 4 values of $\beta$ correspond to Dp-branes with $p = 3, 4, 5, 6$.

We should note, that realistic versions of D-brane inflation models constructed so far have $m \ll 1$ and predict very small $r$, in the range of $10^{-6} - 10^{{-10}}$ \cite{Kachru:2003sx,Martin:2013tda}. This may change with the further development of such models. 
Interestingly, the phenomenological potentials associated with D-brane inflation  \rf{quart}-\rf{lin}, with arbitrary values of $m$ and $r$, also appear in  the theory of pole inflation    \cite{Galante:2014ifa,Terada:2016nqg} describing the cosmological attractors with the pole order $q$ in the kinetic term of the inflaton  field. In this context, $\alpha$-attractors are the pole inflaton models with $q = 2$. The  D-brane inflation potentials  with  $k =4$,  3,  2,  1 corresponding to $\beta=   {5\over 3}, \   {8\over 5}, \  {4\over 3}, \  {3\over 2}$, belong to the class of the pole inflation potentials with $q = {5\over 2},\ {8\over 3}, \  { 3}, \  4 $ respectively, see section \ref{general}.

Unlike  $\alpha$-attractors with $q = 2$, pole inflation for $q\not = 2$ does not  have  deep roots in supergravity. In this respect, it may not have an equally good interpretation in terms of fundamental physics, but such interpretation may be found in the future. At the very least, pole inflation with $q\not = 2$ provides a very powerful tool for development, interpretation and classification of a broad class of cosmological attractors. In particular, it allows to generate all inflaton potentials used in the phenomenological parametrization of inflationary models developed in   \cite{Mukhanov:2013tua,Creminelli:2014nqa,Abazajian:2016yjj}.  This method immediately allows to explain the attractor nature of such potentials, and find many new potentials of desirable type. 

For example, the hilltop inflation potential  $\bigl(1-   {\vp^{4}\over m^{4}} \bigr)^{2}$ \rf{44} and the plateau potential  $\bigl(1-   {m^{4}\over \vp^{4}} \bigr)^{2}$ \rf{quart2} could seem rather {\it ad hoc}. However, in the context of the pole inflation approach, these two models represent the simplest $q = 3/2$ and $q = 5/2$ counterparts of the $q = 2$ $\alpha$-attractor E-model $\bigl(1 - e^{-\sqrt {2\over 3\alpha}\varphi}\bigr)^{2}$ generalizing the Starobinsky model. And even though models with $q = 5/2$ may not have a clear motivation in string theory and supergravity, the ease with which a combination of the simplest T-model  \rf{T} ($q = 2$, yellow lines) and the model  \rf{quart2}  ($q = 5/2$, magenta lines)  cover the sweet spot of the Planck data in Fig. \ref{QQ2} is quite remarkable.

In addition, the pole inflation approach provides a unique way to derive simple general expressions for  $n_{s}$, $r$ and $A_{s}$ \rf{pole}, as well as equation \rf{neqq} for the parameter  $N_{eq}$  introduced in  \cite{Creminelli:2014nqa,Abazajian:2016yjj}, in terms of the residue at the pole $a_{q}$, see section \ref{Char}. We find that the  characteristic scale of inflation $M$ introduced in \cite{Abazajian:2016yjj}, has a particularly simple relation to the residue at the pole, $M^{2} = 4a_{q}$, see \rf{aq}, but we believe that from the point of view of the theory of $\alpha$-attractors it would be more natural to define this scale as $M^{2} = 2a_{q}$ \rf{M3}, or simply as $M^{2} = a_{q}$ \rf{M1}.

 Let us briefly summarize  the main results of our investigation of the inflationary models favored by Planck 2018. We found that some of these models, such as the simplest versions of hilltop inflation and D-brane inflation, are theoretically inconsistent. However, consistent versions of D-brane inflation, in combination with the simplest $\alpha$-attractor models,  can successfully describe most  of the  area in the ($n_{s}$, $r$) space favored by Planck 2018. These two classes of models are complementary to each other:  $\alpha$-attractors  tend to describe the left side of the area in the ($n_{s}$, $r$) space favored by Planck 2018, whereas D-brane models describe the right-hand side of this area, see  Figs. \ref{6pots}, \ref{QQ}, and  \ref{QQ2}. We found that the pole inflation approach to the theory of cosmological attractors can  provide  a unified phenomenological description of the models favored by the Planck 2018, including $\alpha$-attractors and D-brane inflation.

Turning from the investigation of these models  to the future observational missions, it is important to identify a set of specific targets for $r$ and $n_{s}$ to be tested. If B-modes are detected above $r \approx 10^{-2}$, the well motivated models of inflation, such as the monodromy inflation   \cite{Silverstein:2008sg,McAllister:2008hb,McAllister:2014mpa}, may be validated relatively soon. However, if this does not happen, we should clarify what is known about  $r \lesssim 10^{-2}$.  Our conclusions are specific for the range $ 10^{-3} \lesssim r \lesssim 10^{-2}$, where we can present  the B-mode targets valid for   U-duality symmetric  class of $\alpha$-attractors,   and for   $ r \lesssim 10^{-3}$, where we present B-mode targets for which the future precision measurements of $n_s$  will be decisive, see Fig. \ref{7disk2}.

In the  general class of $\alpha$-attractors not related to supergravity, or in the models based on  minimal  $\cN=1$ supergravity,  $\alpha$ is an arbitrary parameter, and $r$ can take any value below the current experimental bound $r\lesssim 6\cdot 10^{-2}$.  The parameter $3\alpha$ is related to the \K\, curvature of the hyperbolic  geometry,  $\mathcal{R}_{{K}}= -{2\over 3\alpha}$. Thus, the search for inflationary B-modes may go beyond investigation of our space-time: It may help us to explore geometry of the internal space of scalar fields responsible for inflation.

In particular, it is possible  that the earliest moments of the Universe are described by maximal supersymmetry theories, including $\cN=8$ supergravity with \E\, U-duality, spontaneously broken to the minimal $\cN=1$ supergravity. If spontaneous symmetry breaking occurs in the potential, then the inflaton kinetic terms - and therefore the potential-independent predictions of $\alpha$-attractors - may reflect the  geometric nature and the symmetries of the original theory  \cite{Ferrara:2016fwe,Kallosh:2017ced}. 

U-duality symmetric  $\alpha$-attractors have $3\alpha=7,6,5,4,3,2,1$, which leads to 7 different predictions for $r$ in the range  $ 10^{-3} \lesssim r \lesssim 10^{-2}$. If B-modes are detected at one of the discrete  levels corresponding to $3\alpha=7,6,5,4,3,2,1$,  as shown in Fig.~\ref{7disk2},  it will provide an evidence for the fundamental structure of the theory  with maximal supersymmetry discussed above.

 Some of these targets may have a different origin. In string theory fibre inflation \cite{Cicoli:2008gp,Kallosh:2017wku} one may encounter $\alpha = 2$. The Starobinsky model~\cite{Starobinsky:1980te}, the Higgs inflationary model~\cite{Salopek:1988qh,Bezrukov:2007ep},  as well as the conformal inflation model~\cite{Kallosh:2013hoa} correspond to $\alpha = 1$.  $\alpha =1/3$ is suggested by the maximal superconformal symmetry  \cite{Kallosh:2019eeu}, see Appendix \ref{conf}.  Yet another target, $\alpha = 1/9$, corresponds to the GL model \cite{Goncharov:1985yu,Linde:2014hfa}.

In addition to a set of targets for $r$, now we have a set of new targets for $n_{s} = 1-\beta/N$, including $\beta=2$ for $\alpha$-attractors, and $\beta={5\over 3},{8\over 5}, {3\over 2},{4\over 3}$ for D-brane inflation. Note that these targets contain some uncertainty. First of all, in some models the  attractor regime $n_{s} = 1-\beta/N$ is reached only for $r \lesssim 10^{{-3}}$. Secondly,  the value of $N$  depends on the process of reheating. However, these general issues can be addressed for each particular model.  With the expected improvement of  precision of determination of $n_s$, the possibility to distinguish various classes of models from each other by comparing  their predictions of $n_{s}$ becomes most interesting and informative \cite{Hardwick:2018zry}. This may become especially relevant in the context of the cosmological attractors discussed in our paper, where the predictions of $n_{s}$ in the small $r$ limit become tightly confined within each $\beta$ stripe.

 In particular, as we already mentioned, $\alpha$-attractors (pole inflation with $q = 2$) tend to describe the  left hand side of the area in the ($n_{s}$, $r$) space favored by Planck 2018, whereas D-brane models and pole inflation with $q > 2$  describe the right hand side of this area,  see  Figs. \ref{6pots}, \ref{QQ}, and  \ref{QQ2}. Therefore even a modest  increase of precision in measurement of $n_{s}$ may provide crucial evidence supporting one of these classes of models.

 \noindent{\bf {Acknowledgments:}} We are grateful to Z. Ahmed, F. Finelli, R. Flauger, M. Hazumi, S. Kachru, L. Knox, Chao-Lin Kuo, J. Martin, L. Page, D. Roest, L. Senatore, E. Silverstein, V. Vennin, R. Wechsler and Y. Yamada    for stimulating discussions.
 This work is supported by SITP and by the US National Science Foundation grant PHY-1720397, by the Simons Foundation Origins of the Universe program (Modern Inflationary Cosmology collaboration), and by the Simons Fellowship in Theoretical Physics.
 
\appendix

 \section{\boldmath Maximal superconformal theory and $3\alpha=1$ benchmark }\label{conf}

 The origin of all 7 benchmarks in Fig.~\ref{7disk2} was explained in detail in    \cite{Ferrara:2016fwe,Kallosh:2017ced}.  It was pointed pout there  that in $\cN=8$ supergravity there is a duality symmetry $E_{7(7)}$, which is  broken to the minimal $\cN=1$ supersymmetry, and the corresponding subgroup of duality is  $E_{7(7)} \mathbb{R}\supset [SL(2, \mathbb{R})]^7$, which describes 7 hyperbolic disks. 
The corresponding kinetic terms are shown in eq. \rf{Lkin}. When all but one disk coordinates are frozen dynamically, one is left with a single disk geometry with $SL(2, \mathbb{R})$ symmetry, which is isomorphic to $SU(1,1)$ symmetry. In such case, the scalar fields are are coordinates of the coset space ${SL(2, \mathbb{R})\over U(1)} $, or ${SU(1,1)\over U(1)} $, and
the kinetic term is the one for a single unit size hyperbolic disk
\be
{\cal L}_{\rm kin} = - {\partial Z \partial \bar Z  \over (1-Z\bar Z)^2}= -{\partial T \partial\bar  T\over (T+\bar T)^2} \ .
\label{single} \ee

 \begin{figure}[!h]
\vspace*{-4mm}
\begin{center}
\includegraphics[width=8.6cm]{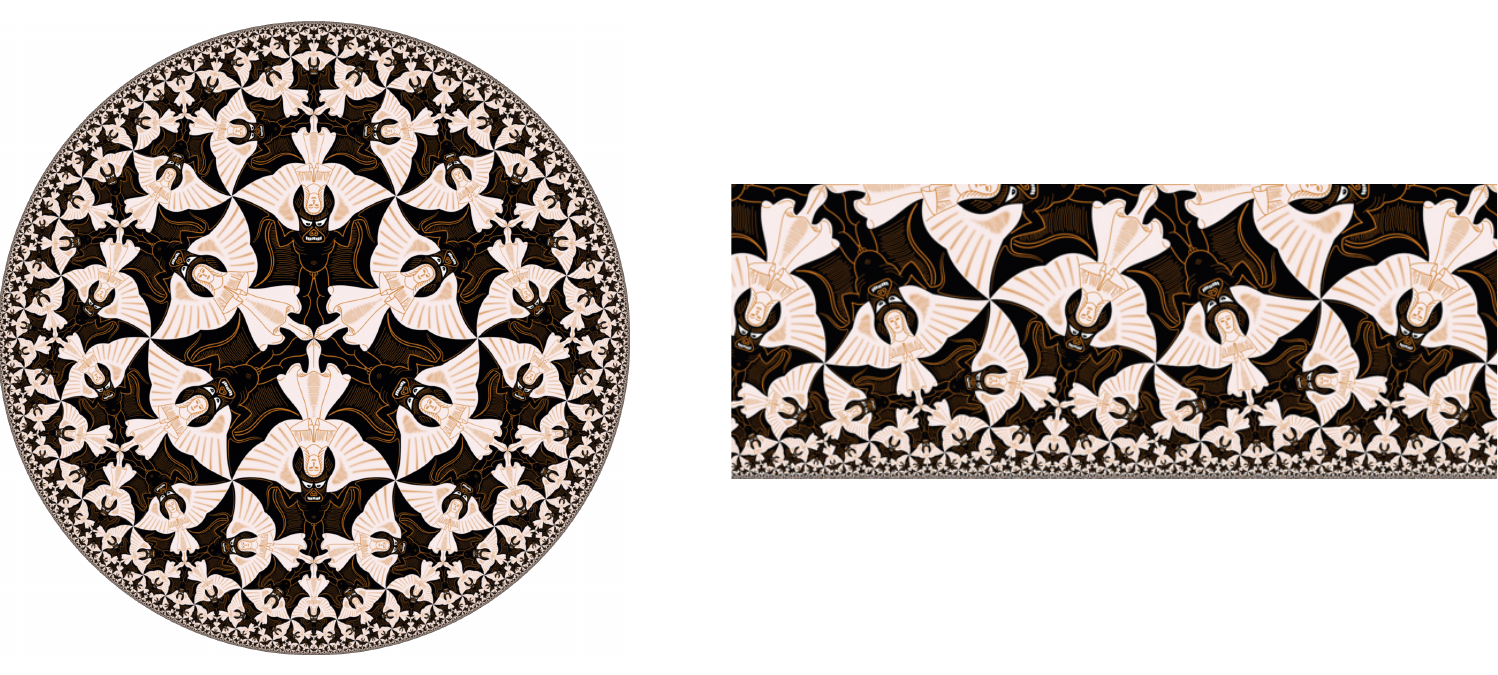}
\caption{\footnotesize On both panels there are Escher's pictures of a Heaven and Hell in disk and half-plane variables, $Z$ and $T$, respectively. The one on the left corresponds to a hyperbolic geometry with the metric in the disk coordinates, as we derive from maximal superconformal theory in eq. \rf{1disk}.
The one on the right corresponds to a hyperbolic geometry with the metric in the half-plane coordinates, as we derive from maximal superconformal theory in eq. \rf{1half}. 
 }
\label{Escher3}
\end{center}
\end{figure}
The relation to maximal superconformal theory of the kinetic term in \rf{single} was pointed out in \cite{Kallosh:2015zsa} with regard to disk and half-plane variables and Escher's pictures in Fig.~\ref{Escher3}. Here we would like to explain this in a more detailed way. 

Extended superconformal theories were studied in \cite{Bergshoeff:1980is,deRoo:1984zyh,Ferrara:2012ui}. The maximal $\cN=4$ superconformal theory has a global duality symmetry $SU(1,1) \times O(6)$. The scalars parametrize the coset space ${SU(1,1)\over U(1)}$. The maximal superconformal theory has a local $SU(4)$ as well as a local $U(1)$ symmetry.  A scalar kinetic term has the form
\be
{1\over 2} D_\mu \phi^\alpha D^\mu \phi_\alpha + hc, \quad \phi^\alpha \phi_\alpha=1, \quad \phi^\alpha = \eta ^{\alpha \beta} \phi_\beta ^* \ ,
\ee
where the scalars are doublets under $SU(1,1)$ symmetry,   $ \alpha=1,2$. It was shown in \cite{Bergshoeff:1980is} that the local $U(1)$  gauge symmetry
can be gauge-fixed by the choice ${\rm Im} \, \phi_1=0$,  so that there is only one independent complex scalar, a coordinate of the unit size Poincar\'e disk, $Z= {\phi_2\over \phi_1}$,
and
\be
{1\over 2} D_\mu \phi_\alpha D^\mu \phi_\alpha + hc\Big |_{{\rm Im} \phi_1=0} = - {\partial Z \partial \bar Z  \over (1-Z\bar Z)^2} \ .
\label{1disk}\ee 
In \cite{Ferrara:2012ui} the choice of the gauge-fixing condition was ${\rm Im} ( \phi_1- \phi_2) =0$,   the independent complex scalar is a coordinate of the half-plane $T=  { \phi_1+ \phi_2\over  \phi_1- \phi_2}$, and the kinetic term is
\be
{1\over 2} D_\mu \phi_\alpha D^\mu \phi_\alpha + hc\Big |_{{\rm Im} (\phi_1- \phi_2) =0} =  -{\partial T \partial\bar  T\over (T+\bar T)^2} \ .
\label{1half}\ee 
There is a simple relation between the disk and half-plane coordinates, known as Cayley relation
\be
Z= {T-1\over T+1}\, , \qquad T= {1+Z\over 1-Z} \ .
\ee
Thus, from the point of view of the maximal superconformal theory, two Escher's pictures in Fig.~\ref{Escher3} correspond to two different choices of gauge-fixing the  local $U(1)$ symmetry. This, in turn, leads to two different coordinate choices for the hyperbolic geometries with the same curvature  $\mathcal{R}= -2$ for the unit size Escher disk with $3\alpha=1$. T-models are simple to formulate  in $Z$-variables with potentials $V_T\sim (Z\bar Z)^n$, E-models are simple in $T$ variables with potentials $V_T\sim (1-T)^{2n}$.

It is important that the maximal $\cN=4$  superconformal theory has a scalar kinetic term  \rf{1disk},  which corresponds to a precise value of the unit size Escher disk.

\bibliography{lindekalloshrefs}
\bibliographystyle{utphys}

\end{document}